\documentclass[a4paper,11pt]{article}

\pdfoutput=1

\usepackage{jcappub}
\usepackage{amsfonts}
\usepackage{mathtools}
\usepackage{bm}
\usepackage{hyperref}

\usepackage{color} 
\definecolor{Blue}{rgb}{0,0,0.9}
\usepackage{soul}

\usepackage[T1]{fontenc}
\title{\boldmath Perturbative Likelihoods for Large-Scale Structure of the Universe}

\author[a,b]{Rodrigo Voivodic}

\affiliation[a]{\small Donostia International Physics Center (DIPC), Paseo Manuel de Lardizabal, 4, 20018 Donostia-San Sebasti\'an, Spain}
\affiliation[b]{\small Instituto de F\'isica Te\'orica (IFT) da Universidade Estadual Paulista (UNESP) and ICTP South American
Institute for Fundamental Research, R. Dr. Bento Teobaldo Ferraz, 271, Bloco II, Barra-Funda - Sao Paulo/SP, Brasil
}

\emailAdd{rodrigo.voivodic@gmail.com}

\abstract{
This work presents a formalism for deriving likelihoods of the cosmological density field directly from first principles within Perturbation Theory (PT). By assuming a perturbative expansion around the Gaussian initial density field and additional stochastic components, we analytically compute two forms of the likelihood. Full marginalization over all underlying fields yields the likelihood of the observed density field, expressed in terms of its summary statistics (such as the power spectrum and bispectrum), which are naturally given by the formalism, and conditioned on model parameters. Marginalizing only over the stochastic fields results in the field-level likelihood. A key strength of this method is its ability to automatically specify the precise combinations of initial field covariances and PT expansion kernels required at each perturbative order (e.g., tree-level power spectrum and bispectrum, and the 1-loop power spectrum). This guarantees that the resulting likelihoods are fully consistent with PT at the chosen order of accuracy, avoiding ad-hoc choices in constructing the statistical model.
}


\begin{document}

\maketitle
\flushbottom

\section{Introduction}
\label{sec:introduction}

Cosmological Perturbation Theory (PT) stands as the primary analytical framework for investigating the large-scale structure (LSS) of the Universe \cite{Bernardeau1, Matsubara1, Matsubara2, Crocce1, Carroll1, Desjacques1}. In contrast to numerical simulations \cite{Angulo1, Vogelsberger1, Takahashi1, Springel1, Teyssier1} and semi-analytical models \cite{Cooray1, Voivodic1, Modi1, Zennaro1, Pellejero1}, PT provides superior control over theoretical predictions and their regimes of validity \cite{Nishimichi1, Carlson1, Konstandin1}. This framework not only predicts observational quantities, such as $n$-point correlation functions \cite{Rubira1, Angulo2, Bertolini1, Carrasco1} and their covariances \cite{Sugiyama1, Wadekar1}, but also facilitates the modeling and marginalization of complex astrophysical processes like the halo-galaxy connection and observational systematics \cite{Mergulhao1, Mergulhao2, Zennaro2}. Consequently, PT has become a cornerstone for constraining cosmological parameters using data from large-scale galaxy surveys \cite{Carrilho1, Zhang1, Damico1}.

Despite its widespread success, the direct application of PT to likelihood computation\footnote{Here, likelihood refers to the probability distribution function of observing a specific dataset conditioned on a set of model parameters.} remains relatively unexplored, with notable exceptions in field-level inference applications \cite{Cabass1, Cabass2, Schmidt1, Schmidt2, Stadler1}. As the crucial statistical link between theoretical predictions and observational data, the likelihood is indispensable for robust cosmological inference. Deriving the likelihood directly from PT principles promises to harness its inherent advantages, ensuring theoretical consistency and affording greater control over the validity range of the entire data analysis pipeline.

In this work, we present a novel methodology for computing the likelihood directly from PT. Our approach begins by equating the observed density field of a given tracer with its theoretical prediction, formulated as a general perturbation expansion up to a specified order. Subsequently, this expression is multiplied by the probability density function (PDF) of the underlying Gaussian primordial fields (which serve as the basis for the perturbative expansion), and these fields are then integrated out. This procedure yields the likelihood of observing a specific density field, conditional upon the covariances of the Gaussian fields and the kernels of the perturbative expansion.

A key advantage of this approach is the inherent self-consistency of the resulting likelihood at the considered perturbative order. This self-consistency arises because, unlike standard cosmological analyses \cite{Damico2, Blot1, DES1}, our method does not require ad-hoc choices for summary statistics, specific sets of theoretical terms, or pre-defined covariance matrices. Furthermore, this approach mitigates the risk of inconsistencies arising from mismatched orders between theoretical predictions and their covariances—a common issue that can inadvertently compromise the theory's range of validity. For example, employing a covariance matrix computed at a lower perturbative order than the mean prediction can artificially restrict the usable range of wavenumbers.

To validate our framework and compare its performance against conventional methods, we contrast our analytically derived likelihoods with commonly used Gaussian likelihood approximations \cite{Ivanov1, Zhang2, DES1, Mergulhao1, Mergulhao2, Rubira1, Voivodic1}. This comparison involves computing the Fisher information and generating numerical posterior distributions for a set of cosmological parameters via the Markov Chain Monte Carlo (MCMC) method.

Our theoretical model employs a perturbative expansion based on two fundamental Gaussian fields: the primordial gravitational potential and a field representing stochastic contributions uncorrelated with this potential. The latter field encapsulates contributions that are not explicitly predictable from the initial conditions, such as those arising from highly non-linear small-scale physics or unresolved galaxy formation processes. While the framework can, in principle, accommodate additional independent Gaussian fields, such extensions are uncommon in existing literature \cite{Desjacques1, Schmidt1, Cabass1} and are not expected to alter the qualitative conclusions of this study.

The remainder of this paper is structured as follows: Sec.~\ref{sec:preliminaries} introduces essential preliminaries, including explicit expressions for the perturbative expansion, likelihood formulation, and prior specifications. Sec.~\ref{sec:first} and \ref{sec:second}, which form the core of this work, detail the computation of the likelihood by marginalizing over all Gaussian fields, up to first and second order in perturbations, respectively. In Sec.~\ref{sec:numerical_posteriors}, we present comparisons of numerical posterior distributions. Sec.~\ref{sec:field} explores field-level likelihoods, derived by marginalizing only over the stochastic field. Finally, Sec.~\ref{sec:conclusions} summarizes our findings and offers concluding remarks. 

\section{Preliminaries}
\label{sec:preliminaries}

When considering local operators in configuration space, the most general density field at order $N$ in the primordial density field and order $M$ in the “primordial” stochastic sector of a tracer $a$ is given by
\begin{align}
    \delta ^{a} _{t}(\textbf{k}|\theta) &= \sum _{n=0}^{N} \sum _{m=0}^{M} (\lambda _{\delta} )^{n}\, (\lambda _{\epsilon})^{m} \, \int_{\mathbb{R}^{3}_{\Lambda_{\delta}}} d^{3} q_{1} \cdots d^{3} q_{n} \int_{\mathbb{R}^{3}_{\Lambda_{\epsilon}}} d^{3} p_{1} \cdots d^{3}p_{m} \, \delta _{D}(\textbf{k} - \textbf{q}_{1 \cdots n} - \textbf{p}_{1 \cdots m})  \nonumber \\
    & \times \mathcal{K}_{(n,m)}^{a}(\textbf{q}_{1}, \cdots, \textbf{q}_{n}, \textbf{p}_{1}, \cdots, \textbf{p}_{m}|\theta) \, \delta (\textbf{q}_{1}) \cdots \delta (\textbf{q}_{n}) \, \epsilon ^{a} (\textbf{p}_{1}) \cdots \epsilon ^{a} (\textbf{p}_{m}) \,,
    \label{eq:theoretical}
\end{align}
where $\mathbb{R}^{3}_{\Lambda} = \left\lbrace \textbf{k} \in \mathbb{R}^{3} \, | \, \textbf{k}^{2} \leq \Lambda^{2} \right\rbrace$ is the space where the fields are integrated, $\theta$ is the set of parameters of the model, $\mathcal{K}_{(n,m)}^{a}(\textbf{q}_{1}, \cdots, \textbf{q}_{n}, \textbf{p}_{1}, \cdots, \textbf{p}_{m}|\theta)$ is a general kernel that encodes the dynamics of the system, $\delta (\textbf{x})$ is the primordial Gaussian density field\footnote{We could also have uses the primordial gravitational potential, which is also Gaussian. In this case, the only differences would be in the explicit functional form of the kernels.}, $\epsilon ^{a}(\textbf{x})$ is the “primordial” Gaussian stochastic field that is different for each tracer, $N$ and $M$ are the orders of the perturbative expansions in both fields, $\Lambda_{\delta}$ and $\Lambda_{\epsilon}$ are the scales where the power spectra of the primordial fields is cutoff, and $\lambda _{\delta}$ and $\lambda _{\epsilon}$ are the perturbative parameters used to keep track of the order of each term and will be set to unity in the end.

The set of parameters $\theta$ is completely general, and its specific definition does not affect any computation of this work. As an example of parameters, we have:
\begin{itemize}
    \item The cosmological parameters ($A_{s}$, $\omega _{m}$, $h$, ...);
    
    \item The redshift ($z$);
    \item The bias parameters ($b_{\mathcal{O}}$);
    \item The parameters of the counter-terms ($c_{s}^{2}$, ...);
    \item The stochastic parameters ($b_{\mathcal{O}'}$);
    \item The non-physical scale used in the computation of the operators ($\Lambda$)\footnote{Note that, for the derivations in this work, it is not important to require the renormalizability of the theory. In principle, our theoretical prediction of the density field might depend on the regularizing scale $\Lambda$}.
\end{itemize}

In the same way, the specific functional form of the kernels does not alter any of the subsequent computations. As examples of kernels and their correspondence with the local operators, we have
\begin{itemize}
    \item $\mathcal{O}^{(0,0)} = \sigma (\Lambda) \Rightarrow \mathcal{K}_{(0,0)}^{a}(\theta) = \sigma^{2}(\Lambda)$. \\The operator with leading order renormalization of $\delta ^{2}(\textbf{x})$;
    
    \item {\small $\mathcal{O}^{(1,0)} = b_{\delta}^{a}(\Lambda) \, \delta(\textbf{x}) + \frac{b^{a}_{\nabla^{2} \delta}(\Lambda)}{k_{\rm NL}^{2}} \, \nabla ^{2}\delta(\textbf{x}) + f_{\rm NL} \, b^{a}_{\phi}(\Lambda) \, \frac{\delta (\textbf{x})}{\nabla^{2}}  \Rightarrow \mathcal{K}_{(1,0)}^{a}(\textbf{k} | \theta) = b_{\delta}^{a}(\Lambda) \, + \, b^{a}_{\nabla^{2} \delta} (\Lambda)\, \left(\frac{k}{k_{\rm NL}}\right)^{2} + \frac{ f_{\rm NL} \, b^{a}_{\phi}(\Lambda)}{k^{2}}$}. \\Some first-order operators on $\delta (\textbf{x})$ parametrizing the leading-order response of the number density of tracers to large-scale perturbations in the primordial density field and potential, as well as the non-local formation of tracers;
    
    \item $\mathcal{O} ^{(0,1)} = b_{\epsilon}^{a}(\Lambda) \, \epsilon(\textbf{x}) + \frac{ b_{\nabla^{2} \epsilon}(\Lambda)}{k_{\rm NL}^{2}} \, \nabla ^{2} \epsilon (\textbf{x}) \Rightarrow\mathcal{K}_{(0,1)}^{a}(\textbf{k} | \theta) = b_{\epsilon}^{a}(\Lambda) + b_{\nabla^{2} \epsilon}(\Lambda) \, \left(\frac{k}{k_{\rm NL}}\right)^{2}$.
    \\Some first-order operators on $\epsilon(\textbf{x})$ parametrizing the leading order corrections in the amplitude and shape of the power spectrum of the stochastic sector. 
    
    \item $\mathcal{O}^{(2,0)} = b_{\delta^{2}}^{a}(\Lambda) \, \delta ^{2}(\textbf{x}) + \delta ^{(2)}(\textbf{x}) \Rightarrow \mathcal{K}_{(2,0)}^{a}(\textbf{k}_{1}, \textbf{k}_{2} | \theta) = b_{\delta^{2}}^{a}(\Lambda) + F_{2}(\textbf{k} _{1}, \textbf{k} _{2})$. \\Some second-order operators on $\delta(\textbf{x})$ parametrizing the second-order response of the number density of tracers to large-scale perturbations in the primordial potential and the non-linear evolution of the density field of matter;
    
    \item $\mathcal{O}^{(1,1)} = b^{a}_{\delta \epsilon}(\Lambda) \, \delta(\textbf{x}) \, \epsilon(\textbf{x})\Rightarrow \mathcal{K}_{(1,1)}^{a}(\textbf{k}_{1}, \textbf{k}_{2} | \theta) = b^{a}_{\delta \epsilon}(\Lambda)$. \\One of the leading-order operators that, non-linearly, generate a correlation between both fields.
\end{itemize}

In this work, we assume uncorrelated primordial density and stochastic fields, both Gaussian with diagonal covariances. The Gaussianity of the primordial density fields is well constrained by data \cite{Leistedt1, Planck1, Shirasak1}, \footnote{Note that, despite our approximation makes impossible to describe the non-linear evolution of non-Gaussian initial conditions, we can describe the responses of the galaxy density fields to large-scale perturbation of the primordial potential through terms with negative power of $k$ in the kernels of Eq.~\eqref{eq:theoretical}.} and, as demonstrated in App.~\ref{app:theory}, we can recover the n-point functions present in the analysis made in the literature through an expansion around a “primordial” Gaussian stochastic field, in the same way it is done for the deterministic part. Another simplification we assume here is the existence of only one “primordial” stochastic field per tracer. This is a usual assumption made in the literature \cite{Mergulhao1, Mergulhao2}, but a larger number of stochastic fields would not change the qualitative results presented here.

Under these simplifications, and taking both cutoff scales to be the same ($\Lambda = \Lambda_{\delta} =\Lambda_{\epsilon}$), the prior for the primordial Gaussian fields can be written as 
{\small \begin{align}
    \mathcal{P}_{\rm prior}[\delta, \left\lbrace \epsilon ^{a} \right\rbrace _{a = 1} ^{\rm T} | \theta] &\propto \left\lbrace \prod _{\textbf{k} \in \mathbb{R}^{3}_{\Lambda}/\mathbb{Z}_{2}} P(\textbf{k}|\theta) \, {\rm det} \left[Q (\textbf{k}|\theta)\right] \right\rbrace ^{-1}  \nonumber \\
    &\times \exp \left( - \int_{\mathbb{R}^{3}_{\Lambda}/\mathbb{Z}_{2}} d^{3}k \, \left\lbrace \frac{| \delta (\textbf{k})|^{2}}{P(\textbf{k} | \theta)} + \sum _{a,b=1} ^{\rm T} \epsilon ^{a} (\textbf{k}) \, \left[ Q ^{-1} (\textbf{k} | \theta) \right] _{ab} \, \bar{\epsilon} ^{b} (\textbf{k}) \right\rbrace \right) \,,
    \label{eq:prior}
\end{align}}
where we dropped the constant normalization factor, $T$ is the number of independent tracers, $\bar{\epsilon}(\textbf{k})$ is the complex conjugate of $\epsilon(\textbf{k})$, $\mathbb{Z}_{2}$ is the group of reflections, $\prod _{\textbf{k} \in \mathbb{R}^{3}_{\Lambda}/\mathbb{Z}_{2}}$ represents a continuous product over the 3D vectors $\textbf{k} \in \mathbb{R}^{3}_{\Lambda}/\mathbb{Z}_{2}$, $P(\textbf{k}|\theta)$ is the primordial power spectrum of the field $\delta(\textbf{k})$, and $\left[ Q(\textbf{k}|\theta) \right]_{ab} = Q_{ab}(\textbf{k}|\theta)$ is the primordial power spectra between the fields $\epsilon^{a}(\textbf{k})$ and $\epsilon^{b}(\textbf{k})$.

In this work, we take the convention that $\left\langle |\delta (\textbf{k})|^{2}\right\rangle = P(\textbf{k}|\theta)$ and $\left\langle |\epsilon(\textbf{k})|^{2}\right\rangle = Q(\textbf{k}|\theta)$. This convention differs by a factor of $2$ from the one assumed in other works \cite{Schmidt1, Schmidt2, Schmidt3, Stadler1}.

As computed in App.~\ref{app:like}, the full likelihood (the one before any marginalization) is given by
\begin{equation}
    \mathcal{P}_{\rm full}[\left\lbrace \delta^{a}_{o} \right\rbrace _{a=1}^{\rm T} | \delta, \left\lbrace \epsilon^{a}  \right\rbrace _{a=1}^{\rm T}, \theta] = \prod _{a=1} ^{\rm T} \left\lbrace \prod _{\textbf{k} \in \mathbb{R}^{3}_{\Lambda _{o}} / \mathbb{Z}_{2}}\delta _{D} \left[ \delta _{o}^{a}(\textbf{k}) - \delta _{t}^{a}(\textbf{k}|\theta) \right] \right\rbrace \,,
    \label{eq:full}
\end{equation}
where $\mathbb{R}^{3}_{\Lambda _{o}}$ is the space of 3D vectors with size smaller than the scale $\Lambda_{o}$\footnote{Also called as $k_{\rm max}$ in the analysis of n-point functions.} where data is given (which might differ from the scale $\Lambda$ used in the regularization of the theoretical predictions \cite{Assassi1}), $\delta _{o}^{a}(\textbf{k})$ is the observed density field of tracer $a$, and $\delta_{D}\left[ f(\textbf{k})\right]$ is the complex evaluated Dirac delta function, defined by
\begin{equation}
    \delta_{D}[f(\textbf{k})] \coloneqq \delta_{D}[f_{r}(\textbf{k})] \, \delta_{D}[f_{i}(\textbf{k})] \,,
\end{equation}
with the subscripts $r$ and $i$ corresponding to the real and imaginary parts of the field, respectively. 

The main goal of the following sections will be to compute the likelihood of the data conditioned only on the parameters (in Secs.~\ref{sec:first} and \ref{sec:second}) and on the parameters and the primordial density field (in Sec.~\ref{sec:field}). This can be done, for the former case, by the integral
\begin{align}
    \mathcal{P}_{\rm like}[\left\lbrace \delta _{o} \right\rbrace _{a=1}^{\rm T}| \theta] &= \int \mathcal{D}\delta_{r} \, \mathcal{D}\delta_{i} \, \left[ \prod _{a=1} ^{\rm T} \mathcal{D}\epsilon_{r}^{a} \, \mathcal{D}\epsilon_{i}^{a} \right] \,  \, \mathcal{P}_{\rm prior}[\delta, \left\lbrace \epsilon ^{a}  \right\rbrace _{a = 1} ^{\rm T} | \theta] \nonumber \\
    &\times \mathcal{P}_{\rm full}[\left\lbrace \delta _{o} \right\rbrace _{a=1}^{\rm T} | \delta (\textbf{k}), \left\lbrace \epsilon^{a} \right\rbrace _{a=1}^{\rm T}, \theta] \,.
    \label{eq:like}
\end{align}

For simplicity, in the rest of this work, we will consider only the case of a single tracer ($T=1$); therefore, we will drop the upper index $a$ from now on. The case of multiple tracers will be studied in future work.

\section{First-order likelihood (\texorpdfstring{$N+M\leq1$}{NM1})}
\label{sec:first}

In this section, we compute the likelihood using only the first-order terms ($N+M\le1$) of Eq.~\eqref{eq:theoretical}. This is the simplest case, but still used in the literature \cite{DES1}. It also provides a direct comparison with the Gaussian likelihood usually used for power spectrum analysis. This comparison is explored numerically in Sec.~\ref{sec:numerical_posteriors}.

We assume that both primordial fields contribute equally to the perturbative expansion, thus setting $\lambda = \lambda _{\delta} = \lambda _{\epsilon}$. This simplification makes computations easier and is sometimes assumed in the literature \cite{Mergulhao1, Mergulhao2}, but again, the general case can be easily read from the results presented here.

To simplify the notation, in this and the next sections, we will consider the duplet field
\begin{equation}
    \phi(\textbf{k}) = \begin{bmatrix}
        \delta (\textbf{k}) \\
        \epsilon(\textbf{k})
    \end{bmatrix} \,,
\end{equation}
and the tensorial kernels 
\begin{align}
    \mathcal{K}_{0}(\theta) &= \mathcal{K}_{(0,0)}(\theta) \,; \nonumber \\
    \mathcal{K}_{1}(\textbf{k}|\theta) &= \begin{bmatrix}
        \mathcal{K}_{(1,0)}(\textbf{k}|\theta)  \\
        \mathcal{K}_{(0,1)}(\textbf{k}|\theta) 
    \end{bmatrix}\,; \nonumber \\ 
    \mathcal{K}_{2}(\textbf{k}_{1}, \textbf{k}_{2}|\theta) &= \begin{bmatrix}
        \mathcal{K}_{(2,0)}(\textbf{k}_{1}, \textbf{k}_{2}|\theta) && \frac{1}{2} \mathcal{K}_{(1,1)}(\textbf{k}_{1}, \textbf{k}_{2}|\theta)\\
        \frac{1}{2}\mathcal{K}_{(1,1)}(\textbf{k}_{1}, \textbf{k}_{2}|\theta) && \mathcal{K}_{(0,2)}(\textbf{k}_{1}, \textbf{k}_{2}|\theta) 
    \end{bmatrix} \,;
    \label{eq:tensorial_kernels}
\end{align}

We will also use Einstein's notation of repeated indices without distinction between upper and lower indices, and define the spaces
\begin{align}
    \mathbb{D} &= \mathbb{R}^{3}_{\Lambda_{o}}/\mathbb{Z}_{2} \,, \\
    \mathbb{G} &= \mathbb{R}^{3}_{\Lambda}/\mathbb{Z}_{2} \,,
\end{align}
as the region where independent data is present, and the Gaussian fields are defined, respectively. 

In this notation, the prior of Eq.~\eqref{eq:prior} becomes
{\small \begin{align}
    \mathcal{P}_{\rm prior}[\phi| \theta] &\propto  \left[ \prod _{\textbf{k} \in \mathbb{G}} {\rm det} \, C(\textbf{k}|\theta)  \right]^{-1} \exp \left\lbrace - \int_{\mathbb{G}} d^{3}k \, \phi _{a}(\textbf{k})\, \left[C^{-1}(\textbf{k}|\theta)\right]^{ab} \, \bar{\phi}_{b}(\textbf{k})\right\rbrace\,,
    \label{eq:prior_phi}
\end{align}}
where
\begin{equation}
    C(\textbf{k}|\theta) = \begin{bmatrix}
        P(\textbf{k}|\theta) && 0 \\
        0 && Q(\textbf{k}|\theta)
    \end{bmatrix}\,.
    \label{eq:linear_covariance}
\end{equation}

\subsection{Likelihood computation}
\label{ssec:computation_first}

At linear order, the full likelihood of Eq.~\eqref{eq:full} can be rewritten as
\begin{align}
    \mathcal{P}^{(0)}_{\rm full}[\delta _{o}| \delta, \epsilon,\theta] &= \prod _{\textbf{k} \in \mathbb{D}}\delta _{D} \left[ \delta _{o}(\textbf{k}) - \delta _{t}(\textbf{k}|\theta) \right] \nonumber  \\
    &= \prod _{\textbf{k} \in \mathbb{D}} \int_{\mathbb{R}^{3}} \frac{d \omega_{r} (\textbf{k})}{2 \pi} \int_{\mathbb{R}^{3}} \frac{d \omega_{i} (\textbf{k})}{2 \pi} \exp \left\lbrace - i \omega (\textbf{k}) \left[ \bar{\delta} _{o}(\textbf{k}) - \bar{\delta} _{t}(\textbf{k}|\theta)\right]\right\rbrace \nonumber \\
    &= \int \mathcal{D}\omega_{r} \, \mathcal{D}\omega_{i} \, \exp \left\lbrace -i \sum_{\textbf{k} \in \mathbb{D}} \, \omega (\textbf{k}) \left[ \bar{\delta} _{o}(\textbf{k}) - \bar{\delta} _{t}(\textbf{k}|\theta) \right] \right\rbrace \nonumber \\
    &= \int \mathcal{D}\omega_{r} \, \mathcal{D}\omega_{i} \, \exp \left\lbrace -i \sum_{\textbf{k} \in \mathbb{D}} \omega (\textbf{k}) \left[ \bar{\delta} _{o}(\textbf{k}) - \lambda \, \mathcal{K}_{1}^{a}(\textbf{k}|\theta) \, \bar{\phi} _{a}(\textbf{k}) \right] \right\rbrace \,,
    \label{eq:linear_full}
\end{align}
where we have defined the path integral measure as usual
\begin{equation}
    \int \mathcal{D} \omega = \int_{\mathbb{R}^{3}} \prod_{\textbf{k} \in \mathbb{D}} \frac{d \omega (\textbf{k})}{2 \pi} \,,
\end{equation}
and the auxiliary field $\omega(\textbf{k})$ respect the reality condition $\bar{\omega}(\textbf{k}) = \omega(-\textbf{k})$, as shown in App.~\ref{app:like}.

The term proportional to $\mathcal{K}_{0}(\theta)$ will not contribute if $\delta _{o}(0) = 0$, which is always the case by construction. Thus, we will not carry this term in the computations.

It is worth pointing out that, despite the integrals on $\textbf{k}$ being limited to $\mathbb{D}$, the path integrals over $\omega(\textbf{k})$ are taken in whole $\mathbb{R}$.

To compute the likelihood of the data given the parameters, we have to marginalize over the primordial Gaussian fields using the prior in Eq.~\eqref{eq:prior_phi}, as shown in Eq.~\eqref{eq:like}. This integral becomes
{\small \begin{align}
    \mathcal{P}^{(0)}_{\rm like}[\delta _{o}|\theta] &\propto \left[ \prod _{\textbf{k} \in \mathbb{G}} {\rm det} \, C(\textbf{k}|\theta) \right]^{-1} \int \mathcal{D}\omega_{r} \, \mathcal{D}\omega_{i} \, \exp \left[ -i \sum_{\textbf{k} \in \mathbb{D}} \omega (\textbf{k}) \, \bar{\delta} _{o}(\textbf{k})\right] \nonumber \\
    &\times \int \mathcal{D}\phi_{r} \, \mathcal{D}\phi_{i} \, \exp \left\lbrace i\lambda\sum_{\textbf{k} \in \mathbb{D}} \omega (\textbf{k}) \, \mathcal{K}_{1}^{a}(\textbf{k}|\theta) \, \bar{\phi}_{a}(\textbf{k}) - \int_{\mathbb{G}} d^{3}k\, \phi _{a}(\textbf{k})\, \left[C^{-1}(\textbf{k}|\theta)\right]^{ab} \, \bar{\phi}_{b}(\textbf{k}) \right\rbrace \nonumber \\
    &\propto \int \mathcal{D}\omega_{r} \, \exp \left\lbrace -\sum_{\textbf{k} \in \mathbb{D}} \left[\frac{\lambda ^{2}}{4V(\textbf{k})} \, \mathcal{K}_{1}^{a}(\textbf{k}|\theta) \, C_{ab}(\textbf{k}|\theta) \, \mathcal{K}_{1}^{b}(\textbf{k}|\theta) \, \omega_{r}^{2}(\textbf{k}) + i \, \omega_{r} (\textbf{k}) \, \delta_{o,r}(\textbf{k}) \right] \right\rbrace \nonumber \\
    &\times (r \rightarrow i) \nonumber \\
    &\propto \left[\prod_{\textbf{k} \in \mathbb{D}} \frac{\lambda ^{2}}{V(\textbf{k})} \, \mathcal{K}_{1}^{a}(\textbf{k}|\theta) \, C_{ab}(\textbf{k}|\theta) \, \mathcal{K}_{1}^{b}(\textbf{k}|\theta) \right]^{-1/2} \, \exp \left\lbrace - \sum_{\textbf{k} \in \mathbb{D}} \frac{V(\textbf{k}) \,\delta_{o,r}^{2}(\textbf{k})}{\lambda^{2} \, \mathcal{K}_{1}^{a}(\textbf{k}|\theta) \, C_{ab}(\textbf{k}|\theta) \, \mathcal{K}_{1}^{b}(\textbf{k}|\theta)}\right\rbrace \nonumber \\ 
    &\times (r \rightarrow i) \nonumber \\
    &\propto \left[ \prod_{\textbf{k} \in \mathbb{D}} \mathcal{K}_{1}^{a}(\textbf{k}|\theta) \, C_{ab}(\textbf{k}|\theta) \, \mathcal{K}_{1}^{b}(\textbf{k}|\theta) \right]^{-1} \, \exp \left\lbrace - \sum _{\textbf{k} \in \mathbb{D}} \frac{ V(\textbf{k}) \, |\delta_{o}(\textbf{k})|^{2}}{\mathcal{K}_{1}^{a}(\textbf{k}|\theta) \, C_{ab}(\textbf{k}|\theta) \, \mathcal{K}_{1}^{b}(\textbf{k}|\theta)}\right\rbrace \,,
    \label{eq:like_first}
\end{align}}
where we have taken $\lambda = 1$ in the last line and dropped the constants of $2\pi$ coming from the Gaussian integrals:
\begin{align}
    &\int \mathcal{D}\phi \exp \left\lbrace \sum_{\textbf{k} \in \mathbb{D}} \left[ -\frac{1}{2} \phi_{a}(\textbf{k}) \, M^{ab}(\textbf{k}) \, \phi_{b}(\textbf{k}) - i \, J^{a}(\textbf{k}) \, \phi _{a}(\textbf{k})\right] \right\rbrace \nonumber \\
    &\propto \left[ \prod _{\textbf{k} \in \mathbb{D}} {\rm det} \, M(\textbf{k})\right] ^{-1/2} \exp \left\lbrace - \frac{1}{2} \sum_{\textbf{k} \in \mathbb{D}} J_{a}(\textbf{k}) \, \left[M^{-1}(\textbf{k})\right]^{ab}\, J_{b}(\textbf{k})\right\rbrace \,,
\end{align}
and we used that
\begin{equation}
    \int_{\mathbb{D}} d^{3}k = \sum _{\textbf{k} \in \mathbb{D}} V(\textbf{k}) \,,
    \label{eq:integral_sum}
\end{equation}
where $V(\textbf{k})$ is the volume of the infinitesimal cell around $\textbf{k}$.

Note that in Eq.~\eqref{eq:like_first}, all modes with $|\textbf{k}| > \Lambda{o}$ of the primordial Gaussian fields are naturally integrated out and do not contribute to the final likelihood. Therefore, we can use any value of $\Lambda$ in the theoretical expansion of Eq.~\eqref{eq:theoretical} and, for the prior of Eq.~\eqref{eq:prior_phi}, we could even use different values of $\Lambda$ in each part, although this is not theoretically motivated. This observation contrasts with the findings using the formalism of \cite{Schmidt3}, where the same cutoff scales must be used to perform the path integrals over the initial conditions. 

It is also worth pointing out the dependence of likelihood on the perturbative parameter $\lambda$. Even though we considered only the terms proportional to $\lambda$ in the perturbative expansion of Eq.~\eqref{eq:theoretical}, we get terms that are non-linear in $\lambda$\footnote{In fact, this is a non-perturbative term in $\lambda$.}. This happens because we can exactly invert the covariance matrix $C(\textbf{k}|\theta)$ that is diagonal in both the indexes $\textbf{k}$ and $a$. This tunneling-like term captures all information of the primordial fields, and it is the leading order contribution for the non-Gaussian case (as we will see in Sec.~\ref{sec:second}). Therefore, it should be included at all orders of perturbations without any expansion. This is the reason we are using the superscript $(0)$ instead of $(2)$, in the likelihood.

\subsection{Discrete approximation}
\label{ssec:discrete_first}

In order to make the result of Eq.~\eqref{eq:like_first} easier to apply to data, we consider a discrete approximation of the integral inside the exponential and the product in the normalization. Breaking the integral and product over $\textbf{k}$ in narrow bins, and assuming the $SO(3)$ symmetry group, we can approximate the primordial power spectra and the kernels as constants inside each bin
\begin{align}
    \sum_{\textbf{k} \in \mathbb{D}} \frac{ V(\textbf{k}) \, |\delta _{o}(\textbf{k})|^{2}}{\mathcal{K}_{1}^{a}(\textbf{k}|\theta) \, C_{ab}(\textbf{k}|\theta) \, \mathcal{K}_{1}^{b}(\textbf{k}|\theta)} &\approx \sum_{i=1}^{N_{\rm bins}} N(k_{i}) \frac{P_{o}(k_{i})}{\mathcal{K}_{1}^{a}(k_{i}|\theta) \, C_{ab}(k_{i}|\theta) \, \mathcal{K}_{1}^{b}(k_{i}|\theta)} \,,  \\
    \prod_{\textbf{k} \in \mathbb{D}} \mathcal{K}_{1}^{a}(\textbf{k}|\theta) \, C_{ab}(\textbf{k}|\theta) \, \mathcal{K}_{1}^{b}(\textbf{k}|\theta) &\approx \prod_{i=1}^{N_{bins}} \left[ \mathcal{K}_{1}^{a}(k_{i}|\theta) \, C_{ab}(k_{i}|\theta) \, \mathcal{K}_{1}^{b}(k_{i}|\theta) \right]^{N(k_{i}) } \,,
    \label{eq:observational_power}
\end{align}
where $N_{\rm bins}$ is the number of radial bins used in the measurements, $N(k_{i}) $ is the number of independent modes in the bin $i$, and $P_{o}(k_{i})$ is the observed power spectrum\footnote{Here, for simplicity, we assumed symmetry under SO(3), however, any other symmetry group could have been assumed only needing to change the definition of the observed power spectrum of Eq.~\eqref{eq:observational_power}.}, in the same bin, defined as
\begin{align}
    P_{o}(k_{i}) &\coloneqq \frac{1}{N(k_{i}) } \sum _{|k| = k_{i} - \Delta k_{i}/2} ^{k_{i} + \Delta k_{i}/2}\, \sum _{\hat{k} \in \mathbb{S}^{2}/\mathbb{Z}_{2}} V(\textbf{k}) \, |\delta _{o}(\textbf{k})|^{2} \nonumber \\
    N(k_{i})  &\coloneqq V \, \sum _{|k| = k_{i} - \Delta k_{i}/2} ^{k_{i} + \Delta k_{i}/2}\, \sum _{\hat{k} \in \mathbb{S}^{2}/\mathbb{Z}_{2}} V(\textbf{k})
    \label{eq:power_observational}
\end{align}
where $\mathbb{S}^{2}$ is the set of points in the 2D unitary spherical shell, $\Delta k_{i}$ is the radial width of the bin $i$, and $V$ is the effective volume where the data was taken.

The reduced space we are using ($\mathbb{D}$ instead of $\mathbb{R}^{3}_{\Lambda{o}}$) becomes evidently important here. Because we are considering only modes in $\mathbb{D}$, the number of modes defined above accounts only for the independent modes and not all modes in the spherical shell. This fact gives the correct Gaussian covariance for the power spectrum, in contrast to a wrong factor of $2$ that would appear if $\mathbb{R}^{3}_{\Lambda{o}}$ were considered. A larger discussion, together with analytical computations, is presented in App.~\ref{app:like}.

From the likelihood Eq.~\eqref{eq:like_first}, in addition to the observed power spectrum, we can also identify the theoretical prediction for the power spectrum at tree-level
\begin{equation}
    P_{t}(k|\theta) \coloneqq \mathcal{K}_{1}^{a}(k|\theta) \, C_{ab}(k|\theta) \, \mathcal{K}_{1}^{b}(k|\theta) = \mathcal{K}_{(1,0)}^{2}(k|\theta) \, P(k|\theta) + \mathcal{K}_{(0,1)}^{2}(k|\theta) \, Q(k|\theta) \,.
    \label{eq:power_theoretical}
\end{equation}

Under this discrete approximation, the likelihood of Eq.~\eqref{eq:like_first} becomes:
\begin{equation}
    \mathcal{P}^{(0), {\rm d}}_{\rm like}[\left\lbrace P_{o} (k_{i}) \right\rbrace _{i=1}^{N_{\rm bins}}| \theta] \propto \left[ \prod _{i=1}^{N_{\rm bins}}  P_{t}(k_{i}|\theta)^{-N(k_{i})} \right] \exp \left[ - \sum _{i=1}^{N_{\rm bins}} N(k_{i}) \frac{P_{o}(k_{i})}{P_{t}(k_{i}|\theta)}\right] \,,
    \label{eq:linear_like_discrete}
\end{equation}
where the superscript “d” points out to the discrete approximation taken.

The likelihood of Eq.~\eqref{eq:linear_like_discrete} is a gamma distribution with shape parameter $\alpha = 1$. This differs from the commonly used Gaussian likelihood for the power spectrum \cite{Blot1, Ivanov1, Nishimichi1, Zhang1}. Because of the non-perturbative nature of this likelihood, we can not directly compare its expansion with the expansion of the Gaussian likelihood. To circumvent it, we compare the Fisher matrix of both distributions, in Sec.~\ref{ssec:fisher_first}, and the numerical posterior, for a small set of parameters, in Sec.~\ref{sec:numerical_posteriors}.

This likelihood was also pointed out to be the correct one for the power spectrum of the temperatures fluctuations in the cosmic microwave background in \cite{Carrasco1}. In the same work, it was analyzed the Fisher matrices for the Gaussian and gamma likelihoods, and showed why the parameters in the covariance of the Gaussian likelihood should be kept fixed in the MCMC analysis. 

Eq.~\eqref{eq:linear_like_discrete} makes clear all we get after the marginalization of the likelihood up to first order in $\lambda$. Firstly, through Eq.~\eqref{eq:power_observational}, it gives us the summary statistics (i.e., compression of the observed density field) we should use to get all information of the density field at this order. Secondly, by Eq.~\eqref{eq:power_theoretical}, it says how to theoretically predict these summary statistics from the initial model [Eq.~\eqref{eq:theoretical}] and the prior [Eq.~\eqref{eq:prior}]. Lastly, using Eq.~\eqref{eq:linear_like_discrete}, it relates both the observational and theoretical summary statistics, implicitly defining not just the covariance but all moments of the likelihood distribution. All these predictions are impossible to get from the central limit theorem or using cosmological perturbation theory only to predict the observables.

The results of this framework will be repeated and become clear in the Sec.~\ref{ssec:discrete_second}, where second-order terms are included, and the final likelihood easily compared with the Gaussian case. A more general discussion is also presented there.

\subsection{Fisher matrix}
\label{ssec:fisher_first}

Taking the log of the discrete likelihood [Eq.~\eqref{eq:linear_like_discrete}], we get
\begin{equation}
    -\ln \mathcal{P}^{(0), \rm d}_{\rm like}[\left\lbrace P_{o} (k_{i}) \right\rbrace _{i=1}^{N_{\rm bins}}| \theta] = \sum _{i=1}^{N_{\rm bins}} N(k_{i})  \left[ \ln P_{t}(k_{i}|\theta) + \frac{P_{o}(k_{i})}{P_{t}(k_{i}|\theta)} \right] \,.
\end{equation}

Therefore, the Fisher matrix is 
\begin{align}
    F_{\alpha \beta}[\theta] &= \left\langle - \frac{\partial ^{2}}{\partial \theta _{\alpha} \partial \theta _{\beta}} \ln \mathcal{P}_{\rm like}[\left\lbrace P_{o} (k_{i}) \right\rbrace _{i=1}^{N_{\rm bins}}, \theta] \right\rangle \nonumber \\
    &=  \left\langle \frac{\partial ^{2}}{\partial \theta _{\alpha} \partial \theta _{\beta}} \sum _{i=1}^{N_{\rm bins}} N(k_{i})  \left[ \ln P_{t}(k_{i}|\theta) + \frac{P_{o} (k_{i})}{P_{t}(k_{i}|\theta)} \right] \right\rangle \nonumber \\
    &=  \left\langle \frac{\partial}{\partial \theta _{\alpha}} \sum _{i=1}^{N_{\rm bins}} N(k_{i})  \left[ \frac{1}{P_{t}(k_{i}|\theta)} - \frac{P_{o} (k_{i})}{P_{t}^{2}(k_{i}|\theta)} \right] \frac{\partial P_{t}(k_{i}|\theta)}{\partial \theta _{\beta}} \right\rangle  \nonumber \\
    &=  \left\langle \sum _{i=1}^{N_{\rm bins}} N(k_{i})  \left\lbrace \left[ \frac{1}{P_{t}(k_{i}| \theta)} - \frac{P_{o} (k_{i})}{P_{t}^{2}(k_{i}|\theta)} \right] \frac{\partial ^{2} P_{t}(k_{i}|\theta)}{\partial \theta _{\alpha} \partial \theta _{\beta}} \right. \right. \nonumber \\
    &\left. \left.+ \left[ 2 \frac{P_{o} (k_{i})}{P_{t}^{3}(k_{i}|\theta)} - \frac{1}{P_{t}^{2}(k_{i}|\theta)} \right] \frac{\partial P_{t}(k_{i}|\theta)}{\partial \theta _{\alpha}} \frac{\partial P_{t}(k_{i}|\theta)}{\partial \theta _{\beta}}\right\rbrace \right\rangle \nonumber \\
    &= \sum _{i=1}^{N_{\rm bins}} \frac{N(k_{i}) }{P_{t}^{2}(k_{i} | \theta)} \frac{\partial P_{t}(k_{i}|\theta)}{\partial \theta _{\alpha}} \frac{\partial P_{t}(k_{i}|\theta)}{\partial \theta _{\beta}}\,,
    \label{eq:linear_fisher}
\end{align}
where the expected values are taken over the discrete likelihood of Eq.~\eqref{eq:linear_like_discrete} that has mean equal to the theoretical power spectrum.

This result is the same one that we get from the Gaussian (in the power spectrum) likelihood. Therefore, we should expect to get similar constraints for the parameters, especially in the limit of large number of modes (large volumes), where the higher moments are suppressed and the central limit theorem is valid. We check this intuition numerically in Sec.~\ref{sec:numerical_posteriors}.

\section{Second-order likelihood (\texorpdfstring{$N+M \leq 2$}{NM2})}
\label{sec:second}

In this section, we consider the first nonlinear corrections in Eq.~\eqref{eq:theoretical}. We will keep considering both fields to contribute equally to perturbation theory ($\lambda_{\delta} = \lambda_{\epsilon}$). Therefore, in this section, we consider the expansion with $N+M \leq 2$.

\subsection{Likelihood computation}
\label{ssec:computation_second}

At second order, the full likelihood of Eq.~\eqref{eq:full} can be rewritten as
{\small \begin{align}
    \mathcal{P}^{(2)}_{\rm full}[\delta _{o}| \delta, \epsilon,\theta] &= \prod _{\textbf{k} \in \mathbb{D}}\delta _{D} \left[ \delta _{o}(\textbf{k}) - \delta _{t}(\textbf{k}|\theta) \right] \nonumber  \\
    &= \prod _{\textbf{k} \in \mathbb{D}} \int_{\mathbb{R}^{3}} \frac{d \omega_{r} (\textbf{k})}{2 \pi} \int_{\mathbb{R}^{3}} \frac{d \omega_{i} (\textbf{k})}{2 \pi} \exp \left\lbrace - i \omega (\textbf{k}) \left[ \bar{\delta} _{o}(\textbf{k}) - \bar{\delta} _{t}(\textbf{k}|\theta)\right]\right\rbrace \nonumber \\
    &= \int \mathcal{D}\omega_{r} \, \mathcal{D}\omega_{i} \, \exp \left\lbrace -i \sum_{\textbf{k} \in \mathbb{D}} \omega (\textbf{k}) \left[ \bar{\delta} _{o}(\textbf{k}) - \bar{\delta} _{t}(\textbf{k}|\theta) \right] \right\rbrace \nonumber \\
    &= \int \mathcal{D}\omega_{r} \, \mathcal{D}\omega_{i} \, \exp \left\lbrace -i \sum_{\textbf{k} \in \mathbb{D}} \, \omega (\textbf{k}) \left[ \bar{\delta} _{o}(\textbf{k}) - \lambda \, \mathcal{K}_{1}^{a}(\textbf{k}|\theta) \, \bar{\phi} _{a}(\textbf{k}) \right] \right\rbrace \nonumber \\
    &\times \exp \left\lbrace i \, \lambda^{2} \, \sum_{\textbf{k} \in \mathbb{D}} \, \omega (\textbf{k}) \int_{\mathbb{R}^{3}_{\Lambda}} d^{3}q_{1} \int_{\mathbb{R}^{3}_{\Lambda}}  d^{3}q_{2} \, \delta_{D}(\textbf{k} - \textbf{q}_{12}) \, \bar{\phi}_{a}(\textbf{q}_{1}) \, \mathcal{K}_{2}^{ab}(\textbf{q}_{1}, \textbf{q}_{2}|\theta) \, \bar{\phi}_{b}(\textbf{q}_{2}) \right\rbrace  \nonumber \\
    &= \int \mathcal{D}\omega_{r} \, \mathcal{D}\omega_{i} \, \exp \left\lbrace -i \sum_{\textbf{k} \in \mathbb{D}} \, \omega (\textbf{k}) \left[ \bar{\delta} _{o}(\textbf{k}) - \lambda \, \mathcal{K}_{1}^{a}(\textbf{k}|\theta) \, \bar{\phi} _{a}(\textbf{k}) \right] \right\rbrace \nonumber \\
    &\times \exp \bigg\lbrace i \, \lambda^{2} \,\sum_{\textbf{k} \in \mathbb{D}} \int_{\mathbb{R}^{3}_{\Lambda}} d^{3}q_{1} \int_{\mathbb{R}^{3}_{\Lambda}}  d^{3}q_{2} \, \delta_{D}(\textbf{k} - \textbf{q}_{12}) \, \mathcal{K}_{2}^{ab}(\textbf{q}_{1}, \textbf{q}_{2}|\theta) \big[ \omega_{r}(\textbf{k}) \, \phi_{a,r}(\textbf{q}_{1}) \,  \phi_{b,r}(\textbf{q}_{2}) \nonumber \\
    &- \omega_{r}(\textbf{k}) \, \phi_{a,i}(\textbf{q}_{1}) \,  \phi_{b,i}(\textbf{q}_{2}) + 2\omega_{i}(\textbf{k}) \, \phi_{a,r}(\textbf{q}_{1}) \,  \phi_{b,i}(\textbf{q}_{2})\big] \bigg\rbrace  \,,
    \label{eq:second_full}
\end{align}}
where $\textbf{q}_{12} = \textbf{q}_{1} + \textbf{q}_{2}$, and we assumed the symmetry of the second-order kernel to wavenumber indices. 

Note that the second-order term introduces a correlation between the real and imaginary parts of the fields. Therefore, we can not integrate in $\phi_{r}$ and $\phi_{i}$ independently, as done for the first-order case. To simplify the notation and the computations, we introduce a quartet field\footnote{It is simply a collection of four scalar fields in a vector-like mathematical object that does not transform as a vector.} that combines the real and imaginary parts. We also need to introduce new kernels and covariance
{ \begin{align}
    \Phi(\textbf{k}) &= \begin{bmatrix}
        \phi_{r}(\textbf{k}) \\
        \phi_{i}(\textbf{k})
    \end{bmatrix} \,; \nonumber \\
    \tilde{\mathcal{K}}_{1}(\textbf{k}|\theta) &= \begin{bmatrix}
        \mathcal{K}_{1}(\textbf{k}|\theta) \, \omega_{r}(\textbf{k})\\
        \mathcal{K}_{1}(\textbf{k}|\theta) \, \omega_{i}(\textbf{k})
    \end{bmatrix} \,; \nonumber \\
    \tilde{\mathcal{K}}_{2}(\textbf{q}_{1}, \textbf{q}_{2}|\textbf{k}, \theta) &= \begin{bmatrix}
        \mathcal{K}_{2}(\textbf{q}_{1}, \textbf{q}_{2}|\theta) \, \omega_{r}(\textbf{k}) && \mathcal{K}_{2}(\textbf{q}_{1}, \textbf{q}_{2}|\theta) \, \omega_{i}(\textbf{k})\\
        \mathcal{K}_{2}(\textbf{q}_{1}, \textbf{q}_{2}|\theta) \, \omega_{i}(\textbf{k}) && -\mathcal{K}_{2}(\textbf{q}_{1}, \textbf{q}_{2}|\theta) \, \omega_{r}(\textbf{k})
    \end{bmatrix} \,; \nonumber  \\
    \tilde{C}(\textbf{k}) &= \begin{bmatrix}
        C(\textbf{k}) && 0 \\
        0 && C(\textbf{k})
    \end{bmatrix} \,.
    \label{eq:Phi}
\end{align}}

Combining Eq.~\eqref{eq:second_full} and Eq.~\eqref{eq:prior_phi}, and using the notation introduced above, we have for the second-order likelihood:
{ \begin{align}
    \mathcal{P}^{(2)}_{\rm like}[\delta _{o}|\theta] &\propto \left[ \prod _{\textbf{k} \in \mathbb{D}} {\rm det} \, C(\textbf{k}|\theta) \right]^{-1} \int \mathcal{D}\omega_{r} \, \mathcal{D}\omega_{i} \, \exp \left[ -i \sum_{\textbf{k} \in \mathbb{D}}  \omega (\textbf{k}) \, \bar{\delta} _{o}(\textbf{k})\right] \nonumber \\
    &\times \int \mathcal{D}\Phi \, \exp \left[ i \, \lambda \, \sum_{\textbf{k} \in \mathbb{D}} \tilde{\mathcal{K}}_{1}^{a}(\textbf{k}|\theta) \, \Phi_{a}(\textbf{k})\right] \nonumber \\
    &\times \exp \Bigg\lbrace - \int_{\mathbb{R}^{3}_{\Lambda}}d^{3}q_{1} \int_{\mathbb{R}^{3}_{\Lambda}} d^{3}q_{2} \, \Phi _{a}(\textbf{q}_{1}) \, \Phi_{b}(\textbf{q}_{2})  \nonumber \\
    &\times \left[ \tilde{C}^{-1}(\textbf{q}_{1}|\theta) \, \delta_{D}(\textbf{q}_{2} - \textbf{q}_{1}) - i \sum_{\textbf{k} \in \mathbb{D}} \lambda ^{2} \, \delta_{D}(\textbf{k} - \textbf{q}_{12}) \, \tilde{\mathcal{K}}_{2}(\textbf{q}_{1},\textbf{q}_{2}|\textbf{k}, \theta) \right]^{ab} \Bigg\rbrace \,,
    \label{eq:second_like_unfinished}
\end{align}}
which, again, is just a Gaussian integral. However, this time, the non-linear covariance is not diagonal in the indices $\textbf{k}$ and  $a$.

The expression inside the last exponential can be rewritten using the relationship of Eq.~\eqref{eq:integral_sum}
{ \begin{align}
    &\int_{\mathbb{R}^{3}_{\Lambda}}d^{3}q_{1} \int_{\mathbb{R}^{3}_{\Lambda}} d^{3}q_{2} \, \Phi _{a}(\textbf{q}_{1}) \, \Phi_{b}(\textbf{q}_{2})  \nonumber \\
    &\times \left[ \tilde{C}^{-1}(\textbf{q}_{1}|\theta) \, \delta_{D}(\textbf{q}_{2} - \textbf{q}_{1}) - i \sum_{\textbf{k} \in \mathbb{D}} \lambda ^{2} \, \delta_{D}(\textbf{k} - \textbf{q}_{12}) \, \tilde{\mathcal{K}}_{2}(\textbf{q}_{1},\textbf{q}_{2}|\textbf{k}, \theta) \right]^{ab}  \nonumber \\
    &= \sum_{\textbf{q}_{1}, \textbf{q}_{2} \in \mathbb{R}^{3}_{\Lambda}}V(\textbf{q}_{1}) \, V(\textbf{q}_{2}) \, \Phi_{a}(\textbf{q}_{1}) \, \Phi_{b}(\textbf{q}_{2}) \nonumber \\
    &\times \left[ C^{-1}(\textbf{q}_{1}) \, \frac{\delta_{K}(\textbf{q}_{2} - \textbf{q}_{1})}{V(\textbf{q}_{1})} - i \frac{\lambda^{2} \, \theta_{H}(\Lambda_{o} - |\textbf{q}_{12}|)}{V(\textbf{q}_{12})} \tilde{\mathcal{K}}_{2}(\textbf{q}_{1}, \textbf{q}_{2}|\textbf{q}_{12})\, \right]^{ab} \,,
\end{align}}
where $\theta_{H}(|\textbf{k}|)$ is the Heaviside function coming from the fact that $|\textbf{q}_{12}| = |\textbf{k}| \leq \Lambda_{o}$, as imposed by the Dirac delta function, and $\delta_{K}(\textbf{k})$ is the Kronecker delta symbol, for continuous variables, defined via
\begin{equation}
    \int d^{3}k \, \delta_{D}(\textbf{k}) = \sum_{\textbf{k}} V(\textbf{k}) \, \frac{\delta_{K}(\textbf{k})}{V(\textbf{k})} = 1 \,.
\end{equation}

Note that now, differently from the first-order case, the determinant of the inverse covariance matrix for $\Lambda_{o} < |\textbf{k}| < \Lambda$ will not be canceled by the normalization of the prior. Therefore, for the corrections of the determinant, we will have to consider the sums over all $\mathbb{R}^{3}_{\Lambda}$. The integrals involving the covariance in the exponential will be only over $\mathbb{D}$ because of the linear term have support only in this space.

Using the results of App.~\ref{app:covariance}, the integral over $\Phi(\textbf{k})$ of Eq.~\eqref{eq:second_like_unfinished} becomes
{ \begin{align}
    \mathcal{P}^{(4)}_{\rm like}[\delta _{o}| \theta] &\approx \int \mathcal{D}\omega_{r} \, \mathcal{D}\omega_{i} \, \Bigg\lbrace1 +  \frac{\lambda^{4}}{2} \sum_{\textbf{q}_{1}, \textbf{q}_{2} \in \mathbb{R}^{3}_{\Lambda}} \frac{V(\textbf{q}_{1} )V(\textbf{q}_{2})}{V^{2}(\textbf{q}_{12})} \, \theta_{H}(\Lambda_{o} - |\textbf{q}_{12}|) \nonumber \\
    &\times \tilde{C}_{ab}(\textbf{q}_{1}|\theta) \, \tilde{\mathcal{K}}^{bc}_{2}(\textbf{q}_{1}, \textbf{q}_{2}|\textbf{q}_{12}, \theta) \, \tilde{C}_{cd}(\textbf{q}_{2}|\theta) \, \tilde{\mathcal{K}}^{da}_{2}(\textbf{q}_{2}, \textbf{q}_{1}|\textbf{q}_{12}, \theta) \Bigg\rbrace^{1/2} \nonumber  \\
    &\times \exp \left\lbrace -i\sum_{\textbf{k}\in \mathbb{D}} \, \left[ \omega_{r}(\textbf{k}) \, \delta_{o,r}(\textbf{k}) + \omega_{i}(\textbf{k}) \, \delta_{o,i}(\textbf{k})\right] \right\rbrace \nonumber \\
    &\times \exp \bigg\lbrace - \sum_{\textbf{k}\in \mathbb{D}} \frac{\lambda^{2}}{4 \, V(\textbf{k})} \tilde{\mathcal{K}}_{1} ^{a} (\textbf{k}|\theta) \, \tilde{C}_{ab}(\textbf{k}|\theta) \, \tilde{\mathcal{K}}_{1}^{b} (\textbf{k}|\theta)  \bigg\rbrace\nonumber \\
    &\times \exp \bigg\lbrace  -i \sum_{\textbf{q}_{1}, \textbf{q}_{2} \in \mathbb{D}} \frac{\lambda^{4}}{4\,V(\textbf{q}_{12})} \theta_{H}(\Lambda_{o} - |\textbf{q}_{12}|) \nonumber \\
    &\times \tilde{\mathcal{K}}^{a}_{1} (\textbf{q}_{1}|\theta) \, \tilde{C}_{ab}(\textbf{q}_{1}|\theta) \, \tilde{\mathcal{K}}^{bc}_{2}(\textbf{q}_{1}, \textbf{q}_{2}|\textbf{q}_{12}, \theta) \, \tilde{C}_{cd}(\textbf{q}_{2}|\theta) \, \tilde{\mathcal{K}}^{d}_{1} (\textbf{q}_{2}|\theta) \bigg\rbrace \,.
\end{align}

Explicitly Putting back the dependency on $\omega$, we get the likelihood at $\lambda^{4}$-order
\begin{align}
    \mathcal{P}^{(4)}_{\rm like}[\delta _{o}| \theta] &\approx\int \mathcal{D}\omega_{r} \, \mathcal{D}\omega_{i} \, \Bigg\lbrace1 +  \frac{\lambda^{4}}{2} \sum_{\textbf{q}_{1}, \textbf{q}_{2} \in \mathbb{R}^{3}_{\Lambda}} \frac{V(\textbf{q}_{1} )V(\textbf{q}_{2})}{V^{2}(\textbf{q}_{12})} \, \theta_{H}(\Lambda_{o} - |\textbf{q}_{12}|) \nonumber \\
    &\times C_{ab}(\textbf{q}_{1}|\theta) \, \mathcal{K}^{bc}_{2}(\textbf{q}_{1}, \textbf{q}_{2}|\theta) \, C_{cd}(\textbf{q}_{2}|\theta) \, \mathcal{K}^{da}_{2}(\textbf{q}_{2}, \textbf{q}_{1}|\theta)\left[ \omega_{r}^{2}(\textbf{q}_{12}) + \omega_{i}^{2}(\textbf{q}_{12}) \right] \Bigg\rbrace\nonumber \\
    &\times \exp \left\lbrace -i\sum_{\textbf{k}\in \mathbb{D}} \left[ \omega_{r}(\textbf{k}) \, \delta_{o,r}(\textbf{k}) + \omega_{i}(\textbf{k}) \, \delta_{o,i}(\textbf{k})\right] \right\rbrace \nonumber  \\
    &\times \exp \bigg\lbrace - \sum_{\textbf{k}\in \mathbb{D}}\frac{\lambda^{2}}{4\, V(\textbf{k})} \mathcal{K}_{1} ^{a} (\textbf{k}|\theta) \, C_{ab}(\textbf{k}|\theta) \, \mathcal{K}_{1}^{b} (\textbf{k}|\theta) \left[ \omega_{r}^{2}(\textbf{k}) + \omega_{i}^{2}(\textbf{k})\right] \nonumber \\
    &\times \exp \bigg\lbrace  -i \sum_{\textbf{q}_{1}, \textbf{q}_{2} \in \mathbb{D}}  \frac{\lambda^{4}}{4\,V(\textbf{q}_{12})}\theta_{H}(\Lambda_{o} - |\textbf{q}_{12}|) \nonumber \\
    &\times \mathcal{K}^{a}_{1} (\textbf{q}_{1}|\theta) \, C_{ab}(\textbf{q}_{1}|\theta) \, \mathcal{K}^{bc}_{2}(\textbf{q}_{1}, \textbf{q}_{2}|\theta) \, C_{cd}(\textbf{q}_{2}|\theta) \, \mathcal{K}^{d}_{1} (\textbf{q}_{2}|\theta) \nonumber  \\
    &\times \left[ \omega_{r}(\textbf{q}_{1}) \, \omega_{r}(\textbf{q}_{2}) \, \omega_{r}(\textbf{q}_{12}) + 2\,\omega_{r}(\textbf{q}_{1}) \, \omega_{i}(\textbf{q}_{2}) \, \omega_{i}(\textbf{q}_{12}) - \omega_{i}(\textbf{q}_{1}) \, \omega_{i}(\textbf{q}_{2}) \, \omega_{r}(\textbf{q}_{12})\right]\bigg\rbrace \,.
    \label{eq:like_second}
\end{align}}

Now, we have a cubic dependence on the field $\omega(\textbf{k})$. Therefore, we can not compute the integral in Eq.~\eqref{eq:like_second} directly; we have to use perturbation theory to get the likelihood at the desired order in $\lambda$.

The main way to compute the likelihood above perturbatively is by adding a new complex duplet source field $J(\textbf{k})$ inside the path integrals. This new field will contribute to the term
\begin{equation}
    \mathcal{P}[J] = \exp \left\lbrace -i \sum_{\textbf{k} \in \mathbb{D}} \left[ \omega_{r} (\textbf{k}) \, J_{r}(\textbf{k}) + \omega_{i} (\textbf{k}) \, J_{i}(\textbf{k})\right] \right\rbrace \,.
    \label{eq:J}
\end{equation}
which is equivalent to shifting the observed field by $J(\textbf{k})$.

Using the term above in Eq.~\eqref{eq:like_second}, we get
{ \begin{align}
    \mathcal{P}^{(4)}_{\rm like}[\delta _{o}| \theta] &\approx \Bigg\lbrace 1 -  \frac{\lambda^{4}}{2} \sum_{\textbf{q}_{1}, \textbf{q}_{2} \in \mathbb{R}^{3}_{\Lambda}} \frac{V(\textbf{q}_{1} )V(\textbf{q}_{2})}{V^{2}(\textbf{q}_{12})} \, \theta_{H}(\Lambda_{o} - |\textbf{q}_{12}|) \nonumber \\
    &\times C_{ab}(\textbf{q}_{1}|\theta) \, \mathcal{K}^{bc}_{2}(\textbf{q}_{1}, \textbf{q}_{2}|\theta) \, C_{cd}(\textbf{q}_{2}|\theta) \, \mathcal{K}^{da}_{2}(\textbf{q}_{2}, \textbf{q}_{1}|\theta)\left[ \frac{\partial^{2}}{\partial J_{r}^{2}(\textbf{q}_{12})} +  \frac{\partial^{2}}{\partial J_{i}^{2}(\textbf{q}_{12})}  \right] \Bigg\rbrace \nonumber \\
    &\times \exp \bigg\lbrace - \sum_{\textbf{q}_{1}, \textbf{q}_{2} \in \mathbb{D}}  \frac{\lambda^{4}}{4\,V(\textbf{q}_{12})}\theta_{H}(\Lambda_{o} - |\textbf{q}_{12}|) \nonumber \\
    &\times \mathcal{K}^{a}_{1} (\textbf{q}_{1}|\theta) \, C_{ab}(\textbf{q}_{1}|\theta) \, \mathcal{K}^{bc}_{2}(\textbf{q}_{1}, \textbf{q}_{2}|\theta) \, C_{cd}(\textbf{q}_{2}|\theta) \, \mathcal{K}^{d}_{1} (\textbf{q}_{2}|\theta) \nonumber \\
    &\times \bigg[\frac{\partial^{3}}{\partial J_{r}(\textbf{q}_{1})\partial J_{r}(\textbf{q}_{2})\partial J_{r}(\textbf{q}_{12})} + 2\frac{\partial^{3}}{\partial J_{r}(\textbf{q}_{1}) \partial J_{i}(\textbf{q}_{2})\partial J_{i}(\textbf{q}_{12})} \nonumber \\
    &- \frac{\partial^{3}}{\partial J_{i}(\textbf{q}_{1})\partial J_{i}(\textbf{q}_{2})\partial J_{r}(\textbf{q}_{12})} \bigg] \bigg\rbrace  \mathcal{P}_{\rm like}^{(0)}[\delta _{o} + J \,| \, \theta] \Bigg|_{J_{r} = J_{i} = 0} \,,
    \label{eq:like_with_derivatives}
\end{align}}
where $\mathcal{P}_{\rm like}^{(0)}[\delta _{o} | \theta]$ is the first order likelihood given by Eq.~\eqref{eq:like_first}, and the partial derivatives are defined such that
\begin{equation}
    \frac{\partial}{\partial J(\textbf{q})} \sum_{\textbf{k}} \phi (\textbf{k}) \, J(\textbf{k}) = \sum_{\textbf{k}} \phi (\textbf{k}) \, \delta_{K}(\textbf{k} - \textbf{q}) = \phi(\textbf{q})\,.
\end{equation}

From the expression of the linear likelihood [Eq.~\eqref{eq:like_first}], we can compute the third order functional derivatives
{\small \begin{align}
    &\frac{\partial^{3}}{\partial J_{a}(\textbf{q}_{1})\partial J_{b}(\textbf{q}_{2})\partial J_{c}(\textbf{q}_{3})} \mathcal{P}_{\rm like}^{0}[\delta _{o}(\textbf{k}) + J(\textbf{k})| \theta]\Bigg|_{J_{r} = J_{i} = 0} = 4 \bigg[ \frac{\delta _{o,a}(\textbf{q}_{1}) \, V(\textbf{q}_{1}) \, V(\textbf{q}_{2})}{P_{t}(\textbf{q}_{1}|\theta) P_{t}(\textbf{q}_{2}|\theta)}\delta _{K}^{bc} \, \delta _{K} (\textbf{q}_{3} - \textbf{q}_{2}) \nonumber \\
    &+ \frac{\delta _{o,b}(\textbf{q}_{2}) \, V(\textbf{q}_{2}) \, V(\textbf{q}_{3})}{P_{t}(\textbf{q}_{2}|\theta) P_{t}(\textbf{q}_{3}|\theta)}\delta _{K}^{ac} \, \delta _{K} (\textbf{q}_{3} - \textbf{q}_{1})+ \frac{\delta _{o,c}(\textbf{q}_{3}) \, V(\textbf{q}_{1}) \, V(\textbf{q}_{3})}{P_{t}(\textbf{q}_{1}|\theta) P_{t}(\textbf{q}_{3}|\theta)}\delta _{K}^{ab} \, \delta _{K} (\textbf{q}_{2} - \textbf{q}_{1})\nonumber \\
    &- 2 \frac{\delta _{o,a}(\textbf{q}_{1}) \delta _{o,b}(\textbf{q}_{2}) \delta _{o,c}(\textbf{q}_{3})}{P_{t}(\textbf{q}_{1}|\theta) P_{t}(\textbf{q}_{2}|\theta) P_{t}(\textbf{q}_{3}|\theta)} \, V(\textbf{q}_{1}) \, V(\textbf{q}_{2}) \, V(\textbf{q}_{3}) \bigg] \mathcal{P}_{\rm like}^{(0)}[\delta _{o}| \theta] \,,
    \label{eq:third_derivative}
\end{align}}
where the subscripts run over real and imaginary.

Therefore, the third-order derivatives of the first-order likelihood become
{ \begin{align}
    &\left[\frac{\partial^{3}}{\partial J_{r}(\textbf{q}_{1})\partial J_{r}(\textbf{q}_{2})\partial J_{r}(\textbf{q}_{12})} + 2\frac{\partial^{3}}{\partial J_{r}(\textbf{q}_{1})\partial J_{i}(\textbf{q}_{2})\partial J_{i}(\textbf{q}_{12})} - \frac{\partial^{3}}{\partial J_{i}(\textbf{q}_{1})\partial J_{i}(\textbf{q}_{2})\partial J_{r}(\textbf{q}_{12})} \right]\nonumber \\
    &\times \frac{1}{4} \, \mathcal{P}_{\rm like}^{0}[\delta _{o} + J \,| \, \theta] \Bigg|_{J_{r} = J_{i} = 0} \nonumber \\
    & = \frac{\delta _{o,r}(\textbf{q}_{1}) \, V(\textbf{q}_{1}) \, V(\textbf{q}_{2})}{P_{t}(\textbf{q}_{1}|\theta) \, P_{t}(\textbf{q}_{2}|\theta)}\delta_{K}(\textbf{q}_{12} - \textbf{q}_{2}) + \frac{\delta _{o,r}(\textbf{q}_{2}) \, V(\textbf{q}_{2}) \, V(\textbf{q}_{12})}{P_{t}(\textbf{q}_{2}|\theta) \, P_{t}(\textbf{q}_{12}|\theta)}\delta_{K}(\textbf{q}_{12} - \textbf{q}_{1}) + \nonumber \\
    &+ \frac{\delta _{o,r}(\textbf{q}_{12}) \, V(\textbf{q}_{1}) \, V(\textbf{q}_{12})}{P_{t}(\textbf{q}_{1}|\theta) \, P_{t}(\textbf{q}_{12}|\theta)}\delta_{K}(\textbf{q}_{2}-\textbf{q}_{1}) +2\frac{\delta _{o,r}(\textbf{q}_{1}) \, V(\textbf{q}_{1}) \, V(\textbf{q}_{2})}{P_{t}(\textbf{q}_{1}|\theta) \, P_{t}(\textbf{q}_{2}|\theta)}\delta_{K}(\textbf{q}_{12} - \textbf{q}_{2}) \nonumber \\
    &- \frac{\delta _{o,r}(\textbf{q}_{12}) \, V(\textbf{q}_{1}) \, V(\textbf{q}_{12})}{P_{t}(\textbf{q}_{1}|\theta) \, P_{t}(\textbf{q}_{12}|\theta)}\delta_{K}(\textbf{q}_{2}- \textbf{q}_{1}) \nonumber \\
    &-2 \frac{\delta_{o,r}(\textbf{q}_{1}) \, \delta_{o,r}(\textbf{q}_{2}) \, \delta_{o,r}(\textbf{q}_{12}) + 2\delta_{o,r}(\textbf{q}_{1}) \, \delta_{o,i}(\textbf{q}_{2}) \, \delta_{o,i}(\textbf{q}_{12}) - \delta_{o,i}(\textbf{q}_{1}) \, \delta_{o,i}(\textbf{q}_{2}) \, \delta_{o,r}(\textbf{q}_{12})}{P_{t}(\textbf{q}_{1}|\theta) \, P_{t}(\textbf{q}_{2}|\theta) \, P_{t}(\textbf{q}_{12}|\theta) \, V^{-1}(\textbf{q}_{1}) \, V^{-1}(\textbf{q}_{2}) \, V^{-1}(\textbf{q}_{12})} \,.
\end{align}}

All terms proportional to the Kronecker delta functions will not contribute because they will generate a term $\delta_{o}(0)$, the mean of the density field, which vanishes by construction.

Now, expanding the exponential in Eq.~\eqref{eq:like_with_derivatives} up to order $\lambda^{4}$, and considering the results above for the third-order derivatives, we get the second-order likelihood:
{\small \begin{align}
    \mathcal{P}^{(4)}_{\rm like}[\delta _{o}| \theta] &\approx \mathcal{P}_{\rm like}^{(0)}[\delta _{o}| \theta]  \bigg\lbrace 1 -  \frac{1}{2} \sum_{\textbf{q}_{1}, \textbf{q}_{2} \in \mathbb{R}^{3}_{\Lambda}} V(\textbf{q}_{1} ) \, V(\textbf{q}_{2}) \, \theta_{H}(\Lambda_{o} - |\textbf{q}_{12}|) \nonumber \\
    &\times C_{ab}(\textbf{q}_{1}|\theta) \, \mathcal{K}^{bc}_{2}(\textbf{q}_{1}, \textbf{q}_{2}|\theta) \, C_{cd}(\textbf{q}_{2}|\theta) \, \mathcal{K}^{da}_{2}(\textbf{q}_{2}, \textbf{q}_{1}|\theta)\, \frac{|\delta_{o}(\textbf{q}_{12})|^{2}}{P_{t}^{2}(\textbf{q}_{12}|\theta)}\nonumber  \\
    &+2 \sum_{(\textbf{q}_{1}, \textbf{q}_{2}) \in \mathbb{D}_{2}} V(\textbf{q}_{1}) \, V(\textbf{q}_{2}) \frac{\mathcal{K}^{a}_{1} (\textbf{q}_{1}|\theta) \, C_{ab}(\textbf{q}_{1}|\theta) \, \mathcal{K}^{bc}_{2}(\textbf{q}_{1}, \textbf{q}_{2}|\theta) \, C_{cd}(\textbf{q}_{2}|\theta) \, \mathcal{K}^{d}_{1} (\textbf{q}_{2}|\theta)}{P_{t}(\textbf{q}_{1}|\theta) \, P_{t}(\textbf{q}_{2}|\theta) \, P_{t}(\textbf{q}_{12}|\theta)}\nonumber \\
    &\times \left[\delta_{o,r}(\textbf{q}_{1}) \, \delta_{o,r}(\textbf{q}_{2}) \, \delta_{o,r}(\textbf{q}_{12}) + 2\delta_{o,r}(\textbf{q}_{1}) \, \delta_{o,i}(\textbf{q}_{2}) \, \delta_{o,i}(\textbf{q}_{12}) - \delta_{o,i}(\textbf{q}_{1}) \, \delta_{o,i}(\textbf{q}_{2}) \, \delta_{o,r}(\textbf{q}_{12})\right] \bigg\rbrace \,,
    \label{eq:like_second_final}
\end{align}}
where we set $\lambda=1$, and defined $\mathbb{D}_{2} = \left\lbrace \textbf{q}_{1}, \textbf{q}_{2} \in \mathbb{D} \, | \, (\textbf{q}_{1} + \textbf{q}_{2})^{2} \leq \Lambda_{o}^{2} \right\rbrace$ as the space allowed by the Heaviside theta function.

From the expression for the likelihood in Eq.~\eqref{eq:like_second_final} we can extract some standard quantities. In the second line, we can identify the numerator as the tree-level theoretical bispectrum
\begin{equation}
    B_{t}(\textbf{q}_{1}, \textbf{q}_{2}|\theta) = \mathcal{K}^{a}_{1} (\textbf{q}_{1}|\theta) \, C_{ab}(\textbf{q}_{1}|\theta) \, \mathcal{K}^{bc}_{2}(\textbf{q}_{1}, \textbf{q}_{2}|\theta) \, C_{cd}(\textbf{q}_{2}|\theta) \, \mathcal{K}^{d}_{1} (\textbf{q}_{2}|\theta) \,,
    \label{eq:theoretical_bi}
\end{equation}
where, if we assume no stochastic field, we get a more familiar expression 
{\small \begin{align}
    B_{t}^{(Q(\textbf{k}) = 0)}(\textbf{q}_{1}, \textbf{q}_{2}|\theta) &= \mathcal{K}_{(1,0)} (\textbf{q}_{1}|\theta) \, P(\textbf{q}_{1}|\theta) \, \mathcal{K}_{(2,0)}(\textbf{q}_{1}, \textbf{q}_{2}|\theta) \, P(\textbf{q}_{2}|\theta) \, \mathcal{K}_{(1,0)} (\textbf{q}_{2}|\theta) \nonumber \\
    &= b_{1}^{2} \, b_{2} \, P(\textbf{q}_{1}|\theta) \, P(\textbf{q}_{2}|\theta) \,,
\end{align}}
where, in the second line, we assumed a simple model of galaxy formation where $\mathcal{K}_{(1,0)}(\textbf{k}|\theta) = b_{1}$ and $\mathcal{K}_{(2,0)}(\textbf{q}_{1}, \textbf{q}_{2}|\theta) = b_{2}$.

Paying attention now to the last line of Eq.~\eqref{eq:like_second_final}, we can identify the observed bispectrum, in the same configuration as the theoretical one, before any integration over the redundant degrees of freedom\footnote{Note that this is not a real bispectrum by the standard definition because it is missing the average over a volume.}
{\small \begin{align}
    B_{o}^{(6d)}(\textbf{q}_{1}, \textbf{q}_{2}) &= {\rm Re}[\delta_{o}(\textbf{q}_{1}) \, \delta_{o}(\textbf{q}_{2}) \, \bar{\delta_{o}}(\textbf{q}_{12})] \nonumber \\
    &=\delta_{o,r}(\textbf{q}_{1}) \, \delta_{o,r}(\textbf{q}_{2}) \, \delta_{o,r}(\textbf{q}_{12}) + 2\delta_{o,r}(\textbf{q}_{1}) \, \delta_{o,i}(\textbf{q}_{2}) \, \delta_{o,i}(\textbf{q}_{12}) - \delta_{o,i}(\textbf{q}_{1}) \, \delta_{o,i}(\textbf{q}_{2}) \, \delta_{o,r}(\textbf{q}_{12}) \,.
\end{align}}

In addition to the bispectrum showing up at second-order, we can also identify the term in the second line as the $(2,2)$ one-loop correction to the power spectrum
\begin{equation}
    P_{t,(2,2)}(\textbf{k}|\theta) = \sum_{\textbf{q}}V(\textbf{q}) \, C_{ab}(\textbf{k} - \textbf{q}|\theta) \, \mathcal{K}^{bc}_{2}(\textbf{k} - \textbf{q},\textbf{k} + \textbf{q}|\theta) \, C_{cd}(\textbf{k} + \textbf{q}|\theta) \, \mathcal{K}^{da}_{2}(\textbf{k} + \textbf{q}, \textbf{k} - \textbf{q}|\theta) \,,
\end{equation}
where we make the change of variables $(\textbf{q}_{1}, \textbf{q}_{2}) \rightarrow (\textbf{k}, \textbf{q}) = (\textbf{q}_{1} + \textbf{q}_{2}, \textbf{q}_{2} - \textbf{q}_{1})/2$.

Therefore, looking to the first-order [Eq.~\eqref{eq:like_first}] and second-order [Eq.~\eqref{eq:like_second_final}] likelihoods, we can see some patterns:
\begin{itemize}
    \item They give us the theoretical expression for the n-point functions we should compute, at that order;
    \item They give us the optimal summary statistics of the density field;
    \item By construction, they have the full information about the connection between the theoretical prediction and data.
\end{itemize}

As shown in App.~\ref{app:covariance}, considering higher-order terms in the perturbative expansion of the density field [Eq.~\eqref{eq:theoretical}] will introduce higher n-point functions with $n \leq N+M+1$. On the other hand, considering higher-order terms in the computation of the inverse covariance and its determinant will introduce loop corrections to the n-point functions already presented in the likelihood. Therefore, we can not push to higher loop corrections without introducing higher n-point functions or using higher n-point functions without including loop corrections, at the same order, in the correlation already present.

It is worth pointing out that the determinant correction in Eq.~\eqref{eq:like_second_final} will use information from modes with $|k| > \Lambda_{o}$ of the initial conditions. However, it will never use information from these modes of the data, as expected from the initial filtering done.

Focusing on the covariance correction term in Eq.~\eqref{eq:like_second_final}, it is important to note that it scales as
\begin{equation}
    \frac{\left\langle \Phi^{(2)}\Phi^{(1)}\Phi^{(1)}\right\rangle \, \left\langle \Phi^{(2)}\Phi^{(1)}\Phi^{(1)} \right\rangle}{\left\langle \Phi^{(3)}\Phi^{(3)} \right\rangle} \propto \lambda ^{2} \,,
\end{equation}
where we took the leading-order term to scale the observational part.

Therefore, in the limit where the perturbative expansion of Eq.~\eqref{eq:theoretical} in valid, this term is (quadratically) smaller than one. This fact is significant because it enforces the likelihood to be positive, which makes it normalizable\footnote{Note that this likelihood is always bounded because of the exponential coming from the non-perturbative term (linear term).} and statistically consistent. 

In contrast to the standard likelihoods used for cosmological inference, this likelihood naturally introduces the $k_{\rm max}$ scale to be used: the one where the perturbative correction to the likelihood becomes unity and the PDF gets negative. In practice, it introduces an extra prior for the parameters. This is similar to the prior introduced by the first order likelihood [Eq.~\eqref{eq:like_first}] that forces the theoretical power spectrum to be positive. This extra prior has a non-negligeble effect, as discussed for the case of PNG parameters in Sec.~\ref{ssec:png_parameters}.

\subsection{Discrete approximation}
\label{ssec:discrete_second}

Similarly to what was done in Sec.~\ref{ssec:discrete_first}, we can make the discrete approximation of the likelihood in Eq.~\eqref{eq:like_second_final}. Considering a $SO(3)$ symmetry\footnote{As before, we could easily use a smaller symmetry group, only changing the number of summations we perform in the definition of the observed bispectrum and power spectrum. For instance, we could consider $SO(2)$ symmetry when redshift space distortions are present.}, we will have three degrees of freedom for the bispectrum [the original six minus three of the dimension of $SO(3)$]. Therefore, we can change from $(\textbf{q}_{1}, \textbf{q}_{2})$ to $(\bm{\chi}, \bm{\zeta})$, where $\bm{\zeta}$ are the degrees of freedom that we will integrate and $\bm{\chi}$ are the degrees of freedom that we will bin\footnote{Usual choices of $\bm{\chi}$ are: $(|\textbf{k}|, |\textbf{q}|, \frac{\textbf{k}\cdot \textbf{q}}{|\textbf{k}| \, |\textbf{q}|})$ and $(|\textbf{k}|, |\textbf{q}|, |\textbf{k} + \textbf{q}|)$.}. Using these new coordinates, we can define the $3D$ observational bispectrum as 
\begin{align}
    B_{o}(\bm{\chi}_{i}) &\coloneqq  \frac{1}{N_{\rm t}(\bm{\chi}_{i})}\sum_{\bm{\chi} =\bm{\chi}_{i}-\frac{\Delta \bm{\chi}_{i}}{2}}^{\bm{\chi}_{i}+\frac{\Delta \bm{\chi}_{i}}{2}} \sum_{\bm{\zeta} \in \mathbb{D}_{\zeta}(\bm{\chi})} V(\bm{\zeta}) \, B_{o}^{(6d)}(\bm{\chi}, \bm{\zeta}) \nonumber \\
    N_{\rm t}(\bm{\chi}_{i}) &\coloneqq V\sum_{\bm{\chi} = \bm{\chi}_{i}-\frac{\Delta \bm{\chi}_{i}}{2}}^{\bm{\chi}_{i}+\frac{\Delta \bm{\chi}_{i}}{2}} \sum_{\bm{\zeta} \in \mathbb{D}_{\zeta}(\bm{\chi})} V(\bm{\zeta}) \,,
\end{align}
where $\mathbb{D}_{\zeta}(\bm{\chi}) = \left\lbrace \bm{\zeta} \in \mathbb{R}^{3}\, | \, \textbf{q}_{1}(\bm{\chi}, \bm{\zeta}) \in \mathbb{D} \, \land \, \textbf{q}_{2}(\bm{\chi}, \bm{\zeta}) \in \mathbb{D} \, \land \, |\textbf{q}_{1} + \textbf{q}_{2}| \leq \Lambda_{o}\right\rbrace$, V is the volume where the data is defined, $\Delta \bm{\chi}_{i}$ is the width of the bin $i$, and $N_{\rm t}(\bm{\chi}_{i})$ is the number of independent triangles in the bin $i$.

Using the definition above of the observational bispectrum together with the definition of the theoretical bispectrum of Eq.~\eqref{eq:theoretical_bi}, we get the discrete approximation of the second-order likelihood of Eq.~\eqref{eq:like_second_final}
{ \begin{align}
    &\mathcal{P}^{(4), {\rm d}}_{\rm like}[\left\lbrace P_{o} (k_{i}) \right\rbrace _{i=1}^{N^{\rm P}_{\rm bins}}, \left\lbrace B_{o} (\bm{\chi}_{i}) \right\rbrace _{i=1}^{N^{\rm B}_{\rm bins}}| \theta] \nonumber \\
    &\approx \mathcal{P}_{\rm like}^{(0), {\rm d}} [\left\lbrace P_{o} (k_{i}) \right\rbrace _{i=1}^{N^{\rm P}_{\rm bins}}| \theta]  \bigg[ 1 -  \sum_{i=1}^{N^{P}_{\rm bins}}N(k_{i}) \, \frac{P_{t, (22)}(k_{i}|\theta) \, P_{o}(k_{i})}{P_{t}^{2}(k_{i}|\theta)} \nonumber \\
    &+2 \sum_{i = 1}^{N^{\rm B}_{\rm bins}}V(\bm{\chi}_{i}) \, N_{\rm t}(\bm{\chi}_{i}) \frac{B_{t}(\bm{\chi}_{i} | \theta) \, B_{o}(\bm{\chi}_{i})}{P_{t}(|\textbf{q}_{1,i}| \,|\theta) \, P_{t}(|\textbf{q}_{2,i}| \,|\theta) \, P_{t}(|\textbf{q}_{12,i}| \, |\theta)} \bigg] \,,
    \label{eq:second_like_discrete}
\end{align}}
where $\mathcal{P}_{\rm like}^{(0), {\rm d}} [\left\lbrace P_{o} (k_{i}) \right\rbrace _{i=1}^{N^{\rm P}_{\rm bins}}| \theta]$ is the discrete approximation at first-order [Eq.~\eqref{eq:linear_like_discrete}], $N^{\rm P}_{\rm bins}$ and $N^{\rm B}_{\rm bins}$ are the number of bins used to measure the power spectrum and the bispectrum, respectively, and we have used the fact that all quantities are invariant by $\mathbb{Z}_{2}$ implying that $\sum_{\textbf{q} \in \mathbb{R}^{3}_{\Lambda}} = 2 \sum_{\textbf{q} \in \mathbb{R}^{3}_{\Lambda}/\mathbb{Z}_{2}}$.

For the numerical implementation of the expression above, it is necessary to compute $|\textbf{q}_{1,i}|$, $|\textbf{q}_{2,i}|$, and $|\textbf{q}_{12}|$ as function of $\bm{\chi}_{i}$, but it is usually trivial by standard choices of the variables $\bm{\chi}$ and $\bm{\zeta}$.

Both sums in Eq.~\eqref{eq:second_like_discrete}, have the form
\begin{equation}
    \sum_{\textbf{k}} \frac{S_{t}(\textbf{k}|\theta) \, S_{o}(\textbf{k})}{\sigma^{2}_{S}(\textbf{k}|\theta)} \,
\end{equation}
where $S(\textbf{k})$ is a summary statistics evaluated in $\textbf{k}$, and $\sigma^{2}_{S}(\textbf{k})$ its variance.

These terms are nothing but the first-order terms that we would get from a Gaussian likelihood for the same summary statistics 
\begin{align}
    \mathcal{P}^{G}_{\rm like}[\left\lbrace S(\textbf{k}) \right\rbrace_{\textbf{k}}|\theta] &\propto \exp \left\lbrace -\frac{1}{2} \sum _{\textbf{k}} \frac{\left[ S_{o}(\textbf{k}) - S_{t}(\textbf{k}|\theta)\right]^{2}}{\sigma^{2}_{S}(\textbf{k}|\theta)}\right\rbrace \nonumber \\
    &\approx \exp \left[ -\frac{1}{2} \sum _{\textbf{k}} \frac{S^{2}_{o}(\textbf{k})
    }{\sigma^{2}_{S}(\textbf{k}|\theta)}\right]\left[ 1 + \sum_{\textbf{k}} \frac{S_{t}(\textbf{k}|\theta) \, S_{o}(\textbf{k})}{\sigma^{2}_{S}(\textbf{k}|\theta)} + \mathcal{O}[S_{t}^{2}]\right] \,.
\end{align}

Therefore, once more, we see that our perturbative likelihood gives the consistent expressions for the observational summary statistics, their theoretical prediction, and the connection between them (e.g., the information encoded in the covariance). 

\section{Numerical posteriors}
\label{sec:numerical_posteriors}

In this section, we compare the perturbative likelihood of Eq.~\eqref{eq:linear_like_discrete} with the usual Gaussian likelihood. To accomplish this, we generate a Gaussian field\footnote{For the generation of the fields we used the open source code \href{https://voivodic.github.io/ExSHalos/}{\texttt{ExSHalos}}, based in the work \cite{Voivodic2}.} for $\delta (\textbf{k})$ and for $\epsilon(\textbf{k})$ and compute the numerical posterior of some parameters using the \href{https://emcee.readthedocs.io/en/stable/}{\texttt{emcee}} code \cite{Foreman1}. 

The Gaussian likelihood, used in the comparisons, is defined by
\begin{equation}
    \mathcal{P}^{\rm{G}, \rm{d}}_{\rm like}[\left\lbrace P_{o} (k_{i}) \right\rbrace _{i=1}^{N^{P}_{\rm bins}}| \theta] \propto \exp \left\lbrace - \frac{1}{2} \sum _{i=1}^{N_{\rm bins}} \left[ \frac{P_{o}(k_{i}) - P_{t}(k_{i}|\theta)}{\sigma (k_{i})}\right]^{2}\right\rbrace \,,
    \label{eq:like_gaussian}
\end{equation}
where the variance was fixed to be the linear prediction $\sigma ^{2}(k_{i}) = P_{t}^{2}(k_{i})/N(k_{i})$.

In Sec.~\ref{ssec:standard_parameters}, we show the comparison between both likelihoods for standard cosmological parameters. On the other hand, in Sec.~\ref{ssec:png_parameters}, we focus only in the leading-order parameter that encapsulates the effect of primordial non-Gaussianity (PNG) in the bias expansion\footnote{In other words, the linear response of the density field to large-scale perturbations in the primordial gravitational potential.}. 

\subsection{Standard parameters}
\label{ssec:standard_parameters}

In Tab.~\ref{tab:params_first}, we show the parameters used in the simulation, their fiducial values, and the priors used during the MCMC sample. We have used wide priors to get informative posteriors even for small boxes.
\begin{table}[ht]
    \centering
    \begin{tabular}{|c|c|c|}
    \hline
         Parameters & Fiducial & Prior \\
         \hline
         \hline
         $A_{s}\, h^{2} \times 10^{9}$ & $\approx0.94$ & $[0.2, 1.5]$ \\
         \hline
         $h$ & $0.67$ & $[0.35, 1.0]$ \\
         \hline
         $\omega _{cdm}$ & $\approx 0.117$ & $[0.02, 0.22]$ \\
         \hline
         $F_{(0,1)}(\textbf{k}|\theta) = c_{1}$ & $1$ & $[0, 3]$ \\
         \hline
    \end{tabular}
    \caption{Parameters used in the MCMC together with their fiducial values, and flat prior range.}
    \label{tab:params_first}
\end{table}

\begin{figure}[ht]
    \centering
    \includegraphics[width=\linewidth]{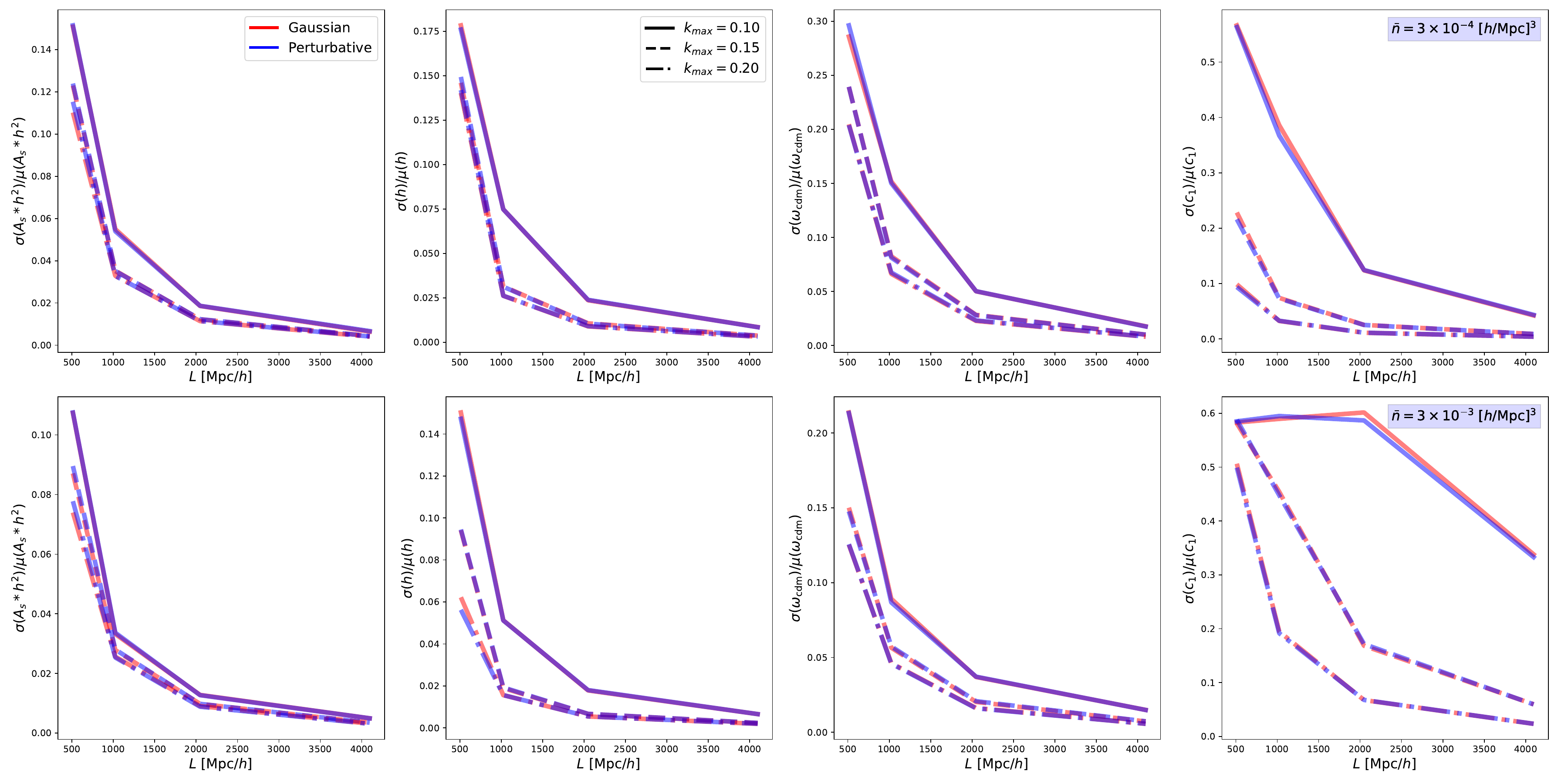}
    \caption{Normalized standard deviations for the three cosmological parameters considered and the stochastic parameter, in the different columns, as a function of the box size. We also plot the results for three different values of $k_{max} = \Lambda_{o}$, in solid, dashed, and dotted-dashed lines for $k_{max} = 0.1$, $0.15$, and $0.2$ $h/$Mpc, respectively. The different rows represent different values used for the constant stochastic power spectrum, with $Q(k) = 10^{4}$ in the first row, and $Q(k) = 10^{3}$ in the second row. The blue lines are the results for the Gaussian likelihood [Eq.~\eqref{eq:like_gaussian}] and the blue lines are the results for the perturbative likelihood [Eq.~\eqref{eq:like_first}].}
    \label{fig:stds_first}
\end{figure}

In Fig.~\ref{fig:stds_first}, we show the normalized standard deviations and their run with the size of the simulated box. There is almost no difference in the information coming from the different likelihoods. We can see small deviations only for the smallest boxes. This conclusion holds for all four parameters considered here.

This finding goes in the direction of the Fisher information matrix of Eq.~\eqref{eq:linear_fisher}, which is the same one for the Gaussian likelihood. The small differences we see in Fig~\ref{fig:stds_first} come from the higher moments (higher derivatives) of the likelihood that are suppressed by the number of modes to $n-1$ for the n-point correlation function, as expected from the central limit theorem.

\begin{figure}[ht]
    \centering
    \includegraphics[width=0.8\linewidth]{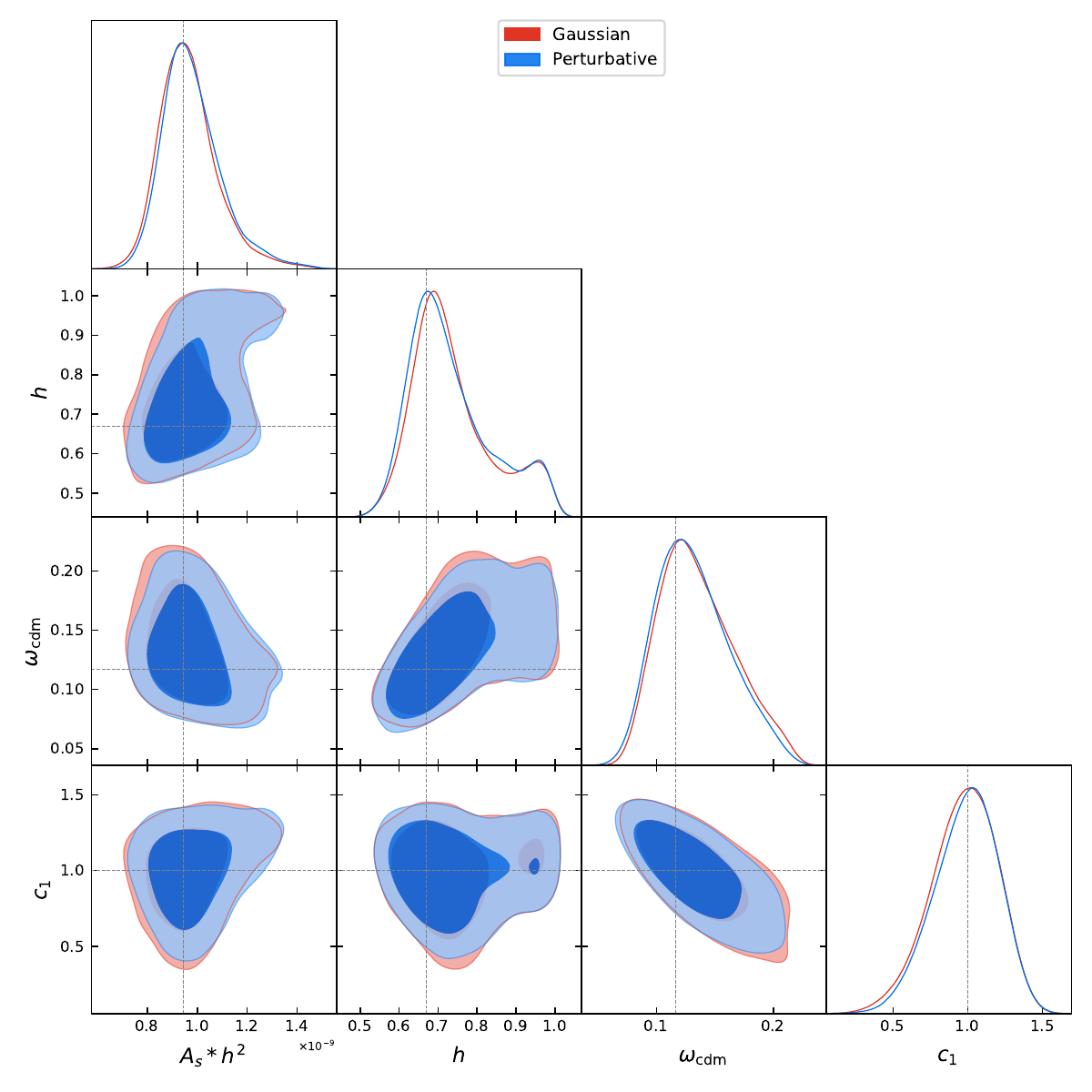}
    \caption{Comparison between the numerical posteriors numerically computed using the Gaussian (red) or the perturbative (blue) likelihoods. The posterior is shown for the smallest box ($L = 512$ Mpc$/h$), where we expect to see more difference between the likelihoods, and for $k_{max} = 0.15$ $h/$Mpc, which is a typical scale used in cosmological analysis with the power spectrum. We also choose $Q(\textbf{k}) = 10^{3}$.}
    \label{fig:chain_first}
\end{figure}

Again, in Fig.~\ref{fig:chain_first}, there is almost no difference between the two posteriors. Even the (possibly numerical) instabilities of the MCMC are correlated in both cases.

\subsection{PNG parameters}
\label{ssec:png_parameters}

In this section, we fix all parameters to be the same as the fiducial ones in the standard analysis, and only vary the leading-order parameter that encodes the effect of primordial non-Gaussianity in the tracer-matter connection. Therefore, we consider the following first order kernel:
\begin{equation}
    \mathcal{K}_{(0,1)}(\textbf{k}|\theta) = 1 + A_{\rm PNG} \left(\frac{k_{\rm PNG}}{k}\right)^{2} \,,
\end{equation}
where $A_{\rm PNG}$ is the PNG parameter\footnote{This parameter is directly related to $b_{\phi} \, f_{\rm NL}$ by a constant that only scales the posteriors and we ignore here.}, and $k_{\rm PNG} = 0.002$ $h/$Mpc is an arbitrary scale selected to keep the parameter dimensionless and order one.

Separately analyzing this PNG parameter is important because it is sensitive to information on very large scales, where there are not many modes and the central limit theorem is not expected to hold. 

For the analysis performed in this section, we fixed $k_{\rm max} = 0.15$ $h/$Mpc, but check that the results stay quantitatively the same for the other values. We also used a flat prior in the interval $[-150, 150]$.

\begin{figure}[ht]
    \centering
    \includegraphics[width=\linewidth]{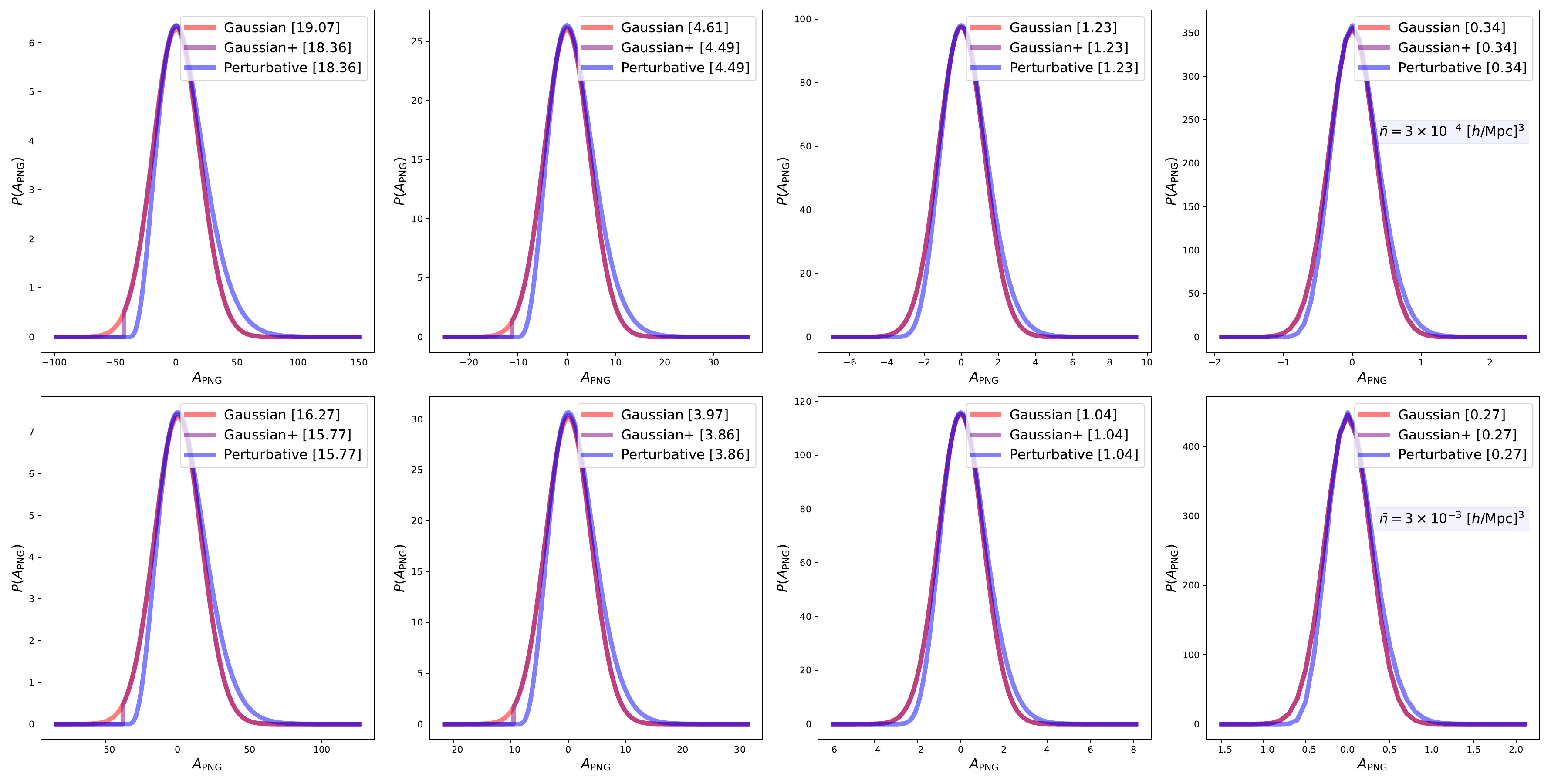}
    \caption{Comparison between the numerical posteriors computed using the Gaussian (red), the Gaussian allowing only positive power spectra (purple), and the perturbative (blue) likelihoods. The values of the standard deviation of each distribution is shown in the parenthesis. The posteriors are shown for $k_{max} = 0.15$ $h/$Mpc. The different rows represent different values used for the constant stochastic power spectrum, with $Q(k) = 10^{4}$ in the first row, and $Q(k) = 10^{3}$ in the second row. Different columns represent different box sizes with $L =$ 512, 1024, 2048, and 4096 Mpc$/h$, from left to right.}
    \label{fig:posteriors_png}
\end{figure}

Fig.~\ref{fig:posteriors_png} shows the comparison between the numerical posteriors. Because our parameter space is now one dimensional, we directly computed the posterior in a regular grid with the same size of the prior and using 3001 points. In addition to the Gaussian and perturbative likelihoods, we also considered the Gaussian case with an extra prior that forces the power spectrum to be positive for all values of k considered. 

In contrast to the results of the last section, the posteriors are not compatible for any box size. This happens because the central limit theorem does not hold when the largest scales contain the majority of the information. On the other hand, we can recover the correct second moment of the distribution when adding an extra prior to the Gaussian likelihood that forces the positiveness of the theoretical power spectrum. This fact highlight an intrinsic problem of using the Gaussian likelihood: it predicts, with non-zero probability, a negative power spectrum. Note that this issue holds for other summary statistics that are physically positive.

For the simple case considered here, the prediction of the standard deviation of the PNG parameter is not very different ($\approx 1 \%$) when comparing the Gaussian and perturbative likelihoods. However, it is not clear what will be the impact of using the wrong likelihood in the case where more parameters are included and higher-order in perturbation theory is considered. 

\section{Field-level likelihoods}
\label{sec:field}

In this section, we study the effect of marginalizing the “full likelihood” [Eq.~\eqref{eq:full}] only with respect to one of the primordial fields. For this work, we consider the marginalization over $\epsilon(\textbf{k})$, because it gives the field-level likelihood that is widely used in the literature \cite{Schmidt1, Schmidt2, Schmidt3, Stadler1}. However, the marginalization only over $\delta(\textbf{k})$ would be very similar and might have some interesting applications in the study of the effects of small-scale physics on the larger scales.

In other words, we are going to compute
\begin{align}
    \mathcal{P}_{\rm like}[\delta _{o} | \delta, \theta] &= \int \mathcal{D}\epsilon_{r} \, \mathcal{D}\epsilon_{i} \, \mathcal{P}_{\rm prior}[\epsilon | \theta] \, \mathcal{P}_{\rm full}[\delta _{o} | \delta , \epsilon, \theta] \,,
    \label{eq:like_field}
\end{align}
where $\mathcal{P}_{\rm prior}[\epsilon | \theta]$ is the prior for the stochastic field alone.

During this section, we will work with the following parts of the perturbative expansion of Eq.~\eqref{eq:theoretical}:
{\begin{align}
    \delta_{t}^{(\leq N, 0)}(\textbf{k}|\delta, \theta) &= \sum_{n=0}^{N} \lambda_{\delta}^{n}\int_{\mathbb{R}^{3}_{\Lambda}} d^{3}q_{1} \cdots d^{3}q_{n} \, \delta_{D}(\textbf{k}- \textbf{q}_{1\cdots n}) \, \mathcal{K}_{(n,0)}(\textbf{q}_{1}, \cdots, \textbf{q}_{n}|\theta) \,; \\
    \delta_{t}^{(0,1)}(\textbf{k}|\theta) &= \lambda_{\epsilon} \, \mathcal{K}_{(0,1)}(\textbf{k}|\theta) \, \epsilon(\textbf{k}) \,; \\
    \delta_{t}^{(1,1)}(\textbf{k}|\theta) &= \lambda_{\delta} \, \lambda_{\epsilon} \int _{\mathbb{R}^{3}_{\Lambda}} d^{3}q \, \mathcal{K}_{(1,1)} (\textbf{k} - \textbf{q}, \textbf{q}|\theta) \, \delta(\textbf{k} - \textbf{q}) \, \epsilon(\textbf{q}) \,; \\
    \delta_{t}^{(0,2)}(\textbf{k}|\theta) &= \lambda_{\epsilon}^{2} \, \int _{\mathbb{R}^{3}_{\Lambda}} d^{3}q_{1} \int _{\mathbb{R}^{3}_{\Lambda}} d^{3}q_{2} \, \delta(\textbf{k} - \textbf{q}_{12}) \, \mathcal{K}_{(0,2)}(\textbf{q}_{1}, \textbf{q}_{2}|\theta) \, \epsilon(\textbf{q}_{1}) \, \epsilon(\textbf{q}_{2}) \,,
    \label{eq:field_terms}
\end{align}}
where we explicitly keep the $\delta$-dependency of $\delta_{t}^{(\leq N, 0)}(\textbf{k}|\delta, \theta)$ to highlight that it is a complicated function computed from the primordial density field.

In the next subsections, we will investigate the effect of each of the terms above on the field-level likelihood. However, we can say what to expect from each term:
\begin{itemize}
    \item The first term is a constant (in $\epsilon$). Thus, it is expected to only shift the observed density field, similarly to the current introduced in Eq.~\eqref{eq:J}. This term contains the information of all correlation functions of the deterministic part;
    \item The second term introduces a decoupled stochastic field. It will introduce noise in the theoretical power spectrum, generating a contribution to the power spectrum $\propto Q(\textbf{k})$;
    \item The third term introduces the coupling between the deterministic and stochastic field, at leading order. We expect it to contribute to the total bispectrum, generating something $\propto P(\textbf{k}) \, Q(\textbf{k})$;
    \item The fourth term introduces the first non-Gaussian correction to the stochastic field. It will introduce noise in the theoretical bispectrum, generating a term $\propto Q^{2}(\textbf{k})$.
\end{itemize}

All computations in this section are similar to the ones in Sec.~\ref{ssec:computation_first} and Sec.~\ref{ssec:computation_second}. Therefore, we will not give as many details as in the former sections.

\subsection{\texorpdfstring{$(0,1)$}{01}-term}
\label{ssec:field_01}

We start with the simplest case where we consider only the $\delta_{t}^{(\leq N, 0)}(\textbf{k}|\delta, \theta)$ term, which is a constant in $\epsilon$, and $\delta_{t}^{(0,1)}(\textbf{k}|\theta)$, that does not introduce any convolution.

For this simple case, the likelihood will not capture any information about the bispectrum of the stochastic field nor any of its correlations with the deterministic one. However, it will still capture all the n-point functions of the deterministic sector. This might be a good approximation if the perturbative expansion is more sensitive to the order in $\lambda_{\delta}$ than $\lambda_{\epsilon}$. Note, however, that we do not have any \textit{a priori} reason to believe that it is the case in nature \cite{Mergulhao1, Mergulhao2, Zennaro1, Zhang1, Zhang2}.

Under these simplifications, the \textit{full likelihood} becomes
\begin{align}
    \mathcal{P}^{(0,1)}_{\rm full}[\delta _{o}| \delta, \epsilon,\theta] &= \prod _{\textbf{k} \in \mathbb{D}}\delta _{D} \left[ \delta _{o}(\textbf{k}) - \delta _{t}(\textbf{k}|\theta) \right] \nonumber  \\
    &= \prod _{\textbf{k} \in \mathbb{D}} \int_{\mathbb{R}^{3}} \frac{d \omega_{r} (\textbf{k})}{2 \pi} \int_{\mathbb{R}^{3}} \frac{d \omega_{i} (\textbf{k})}{2 \pi} \exp \left\lbrace - i \omega (\textbf{k}) \left[ \bar{\delta} _{o}(\textbf{k}) - \bar{\delta} _{t}(\textbf{k}|\theta)\right]\right\rbrace \nonumber \\
    &= \int \mathcal{D}\omega_{r} \, \mathcal{D}\omega_{i} \, \exp \left\lbrace -i \sum_{\textbf{k} \in \mathbb{D}} \, \omega (\textbf{k}) \left[ \bar{\delta} _{o}(\textbf{k}) - \bar{\delta} _{t}(\textbf{k}|\theta) \right] \right\rbrace \nonumber \\
    &= \int \mathcal{D}\omega_{r} \, \mathcal{D}\omega_{i} \, \exp \bigg\lbrace -i \sum_{\textbf{k} \in \mathbb{D}} \omega (\textbf{k}) \left[ \bar{\delta} _{o}(\textbf{k}) - \bar{\delta}_{t}^{(\leq N, 0)}(\textbf{k}|\delta, \theta)  - \delta_{t}^{(0,1)}(\textbf{k}|\theta) \right] \bigg\rbrace \nonumber \\
    &= \int \mathcal{D}\omega_{r} \, \mathcal{D}\omega_{i} \, \exp \bigg\lbrace -i \sum_{\textbf{k} \in \mathbb{D}} \omega (\textbf{k}) \left[ \bar{\delta} _{o}(\textbf{k}) - \bar{\delta}_{t}^{(\leq N, 0)}(\textbf{k}|\delta, \theta) \right] \bigg\rbrace \nonumber \\
    &\times \exp \bigg\lbrace i \sum_{\textbf{k} \in \mathbb{D}} \lambda_{\epsilon} \, \omega (\textbf{k}) \,\mathcal{K}_{(0,1)}(\textbf{k}|\theta) \, \bar{\epsilon}(\textbf{k})\bigg\rbrace \,.
\end{align}

After marginalizing over $\epsilon$, the likelihood reduces to
\begin{align}
    \mathcal{P}^{(0,1)}_{\rm like}[\delta _{o}| \delta, \theta] &=\int \mathcal{D}\omega_{r} \, \mathcal{D}\omega_{i} \, \exp \bigg\lbrace -i \sum_{\textbf{k} \in \mathbb{D}} \omega (\textbf{k}) \left[ \bar{\delta} _{o}(\textbf{k}) - \bar{\delta}_{t}^{(\leq N, 0)}(\textbf{k}|\delta, \theta)\right] \bigg\rbrace \nonumber \\ 
    &\times \exp \bigg\lbrace - \frac{\lambda_{\epsilon}^{2}}{4}\sum_{\textbf{k} \in \mathbb{D}} \frac{\mathcal{K}^{2}_{(0,1)}(\textbf{k}|\theta) \, Q(\textbf{k}|\theta)}{V(\textbf{k})} \, \left[\omega_{r}^{2}(\textbf{k})  + \omega_{i}^{2}(\textbf{k})\right]\bigg\rbrace \nonumber \\
    &\propto \left[ \prod_{\textbf{k} \in \mathbb{D}} \mathcal{K}^{2}_{(0,1)}(\textbf{k}|\theta) \, Q(\textbf{k}|\theta) \right]^{-1} \exp \left[ - \sum_{\textbf{k} \in \mathbb{D}}  V(\textbf{k}) \, \frac{|\delta_{o}(\textbf{k}) - \delta_{t}^{(\leq N, 0)}(\textbf{k}|\delta, \theta)|^{2}}{\mathcal{K}^{2}_{(0,1)}(\textbf{k}|\theta) \, Q(\textbf{k}|\theta)}\right] \,,
    \label{eq:like_field_01}
\end{align}
where $\lambda_{\epsilon}$ was set to unity in the last line.

The likelihood above is equal\footnote{Note that we have an extra factor of two both in the power of the normalization and the exponential. This factor comes from the fact that we consider only independent modes and not the whole $\mathbb{R}^{3}_{\Lambda_{o}}$.} to the EFT-likelihood \cite{Schmidt1, Cabass1}. It describes the field likelihood assuming a perfectly Gaussian stochastic field without any correlation with the deterministic part. This simplification fully captures all the information in the power spectrum but fails to capture the information in the higher n-point function, being incomplete\footnote{This likelihood will only have information about the n-point functions of the deterministic sector isolated, for $(n>2)$-point function. However, it has information about all n-point functions of the deterministic sector of the theory} for $n>2$. 

In the same fashion as the likelihood of Eq.~\eqref{eq:like_first}, it is non-perturbative in $\lambda_{\epsilon}$. Therefore, this expression needs to be fully considered at any order (without expanding it) and can not be directly compared with, e.g., the Gaussian likelihood.

\subsection{\texorpdfstring{$(1,1)$}{11}-term}
\label{ssec:field_11}

In this section, we study the effect of the first term that correlates the deterministic and stochastic sectors. This term is still linear in $\epsilon$ but, because of the convolution it introduces, the inversion of the covariance matrix is now non-trivial. As discussed above, we have to consider the terms of the last section because of its non-perturbative nature in Eq.~\eqref{eq:like_field_01}.

Now, the \textit{full likelihood} takes the form
\begin{align}
    \mathcal{P}^{(1,1)}_{\rm full}[\delta _{o}| \delta, \epsilon,\theta] &=  \int \mathcal{D}\omega_{r} \, \mathcal{D}\omega_{i} \, \exp \bigg\lbrace -i \sum_{\textbf{k} \in \mathbb{D}} \omega (\textbf{k}) \left[ \bar{\delta} _{o}(\textbf{k}) - \bar{\delta}_{t}^{(\leq N, 0)}(\textbf{k}|\delta, \theta) \right] \bigg\rbrace \nonumber \\
    &\times \exp \bigg\lbrace i \,\lambda_{\epsilon} \,\sum_{\textbf{k} \in \mathbb{D}}  \omega (\textbf{k}) \int _{\mathbb{R}^{3}_{\Lambda}} d^{3}q\Big[ \,\mathcal{K}_{(0,1)}(\textbf{q}|\theta) \, \delta_{D}(\textbf{k} - \textbf{q}) \nonumber \\
    &+ \lambda_{\delta} \, \mathcal{K}_{(1,1)} (\textbf{k} - \textbf{q}, \textbf{q}|\theta) \, \bar{\delta}(\textbf{k} - \textbf{q}) \Big] \bar{\epsilon}(\textbf{q})\bigg\rbrace \,.
    \label{eq:full_field_11}
\end{align}

Then, performing the marginalization over $\epsilon$ we get
\begin{align}
    \mathcal{P}^{(1,1)}_{\rm like}[\delta _{o}| \delta, \theta] &=\int \mathcal{D}\omega_{r} \, \mathcal{D}\omega_{i} \, \exp \bigg\lbrace -i \sum_{\textbf{k} \in \mathbb{D}} \omega (\textbf{k}) \left[ \bar{\delta} _{o}(\textbf{k}) - \bar{\delta}_{t}^{(\leq N, 0)}(\textbf{k}|\delta, \theta)\right] \bigg\rbrace \nonumber \\ 
    &\times \exp \bigg\lbrace - \frac{\lambda_{\epsilon}^{2}}{4}\sum_{\textbf{k} \in \mathbb{D}} Q(\textbf{k}|\theta) \,V(\textbf{k}) \, \left[\Omega_{r}^{2}(\textbf{k}|\theta)  + \Omega_{i}^{2}(\textbf{k}|\theta)\right]\bigg\rbrace \,,
    \label{eq:like_field_11}
\end{align}
where the new function $\Omega_{r/i}(\textbf{k})$ are defined by:
{\small \begin{align}
    \Omega_{r}(\textbf{k}|\theta) &= \frac{\omega_{r}(\textbf{k}) \, \mathcal{K}_{(0,1)}(\textbf{k}|\theta)}{V(\textbf{k})} + \lambda_{\delta}\sum_{\textbf{q} \in \mathbb{R}_{\Lambda}^{3}} \mathcal{K}_{(1,1)}(\textbf{q} - \textbf{k}, \textbf{q}|\theta) \, \left[ \omega_{r}(\textbf{q}) \, \delta_{r}(\textbf{q}- \textbf{k}) + \omega_{i}(\textbf{q}) \, \delta_{i}(\textbf{q}- \textbf{k})\right] \,; \\
    \Omega_{i}(\textbf{k}|\theta) &= \frac{\omega_{i}(\textbf{k}) \, \mathcal{K}_{(0,1)}(\textbf{k}|\theta)}{V(\textbf{k})} + \lambda_{\delta}\sum_{\textbf{q} \in \mathbb{R}_{\Lambda}^{3}} \mathcal{K}_{(1,1)}(\textbf{q} - \textbf{k}, \textbf{q}|\theta) \, \left[ \omega_{i}(\textbf{q}) \, \delta_{r}(\textbf{q}- \textbf{k}) - \omega_{r}(\textbf{q}) \, \delta_{i}(\textbf{q}- \textbf{k})\right]     \,.
\end{align}}

Collecting only terms up to order $\lambda_{\delta}$, we can rewrite the expression in the brackets above as
\begin{align}
    \Omega_{r}^{2}(\textbf{k}|\theta)  + \Omega_{i}^{2}(\textbf{k}|\theta) &\approx \left[ \frac{\mathcal{K}_{(0,1)}(\textbf{k}|\theta)}{V(\textbf{k})}\right]^{2}\left[ \omega_{r}^{2}(\textbf{k}) + \omega_{i}^{2}(\textbf{k})\right] \nonumber \\
    &+ \frac{2 \, \lambda_{\delta} \, \mathcal{K}_{(0,1)}(\textbf{k}|\theta)}{V(\textbf{k})} \sum_{\textbf{q} \in \mathbb{R}_{\Lambda}^{3}} \mathcal{K}_{(1,1)}(\textbf{q} - \textbf{k}, \textbf{q}|\theta) \nonumber \\
    &\times \begin{bmatrix}
        \omega_{r}(\textbf{k}) & \omega_{i}(\textbf{k})
    \end{bmatrix}
    \begin{bmatrix}
        \delta_{r}(\textbf{q} - \textbf{k}) & \delta_{i}(\textbf{q} - \textbf{k}) \\
        -\delta_{i}(\textbf{q} - \textbf{k}) & \delta_{r}(\textbf{q} - \textbf{k})
    \end{bmatrix}
    \begin{bmatrix}
        \omega_{r}(\textbf{q}) \\
        \omega_{i}(\textbf{q})
    \end{bmatrix} \,.
\end{align}

Therefore, we can identify the covariance matrix as
\begin{align}
    M(\textbf{k}_{1}, \textbf{k}_{2}|\theta) &= Q(\textbf{k}_{1}|\theta) \Bigg\lbrace \frac{\mathcal{K}^{2}_{(0,1)}(\textbf{k}_{1}|\theta)}{V(\textbf{k}_{1})} \delta_{K}(\textbf{k}_{2} - \textbf{k}_{1}) \, \mathbb{I} \nonumber \\
    &+ 2 \, \lambda_{\delta} \, \mathcal{K}_{(0,1)}(\textbf{k}_{1}|\theta) \, \mathcal{K}_{(1,1)}(\textbf{k}_{2} - \textbf{k}_{1}, \textbf{k}_{2}|\theta)
    \begin{bmatrix}
        \delta_{r}(\textbf{k}_{2} - \textbf{k}_{1}) & \delta_{i}(\textbf{k}_{2} - \textbf{k}_{1}) \\
        -\delta_{i}(\textbf{k}_{2} - \textbf{k}_{1}) & \delta_{r}(\textbf{k}_{2} - \textbf{k}_{1})
    \end{bmatrix} \Bigg\rbrace \,,
\end{align}
where the modes with $\Lambda_{o} \leq |\textbf{k}_{2}| < \Lambda$ can be integrated out, leading to a term that is proportional to the average value of the theoretical density field and, therefore, does not contribute.

At leading order, and following similarly to App.~\ref{app:covariance}, we have for the inverse covariance
\begin{align}
    M^{-1}(\textbf{k}_{1}, \textbf{k}_{2}|\theta) &= \frac{V(\textbf{k})}{\mathcal{K}^{2}_{(0,1)}(\textbf{k}_{1}|\theta) \,Q(\textbf{k}_{1}|\theta)} \delta_{K}(\textbf{k}_{2} - \textbf{k}_{1}) \, \mathbb{I} \nonumber \\
    &- \bigg\lbrace \mathcal{K}_{(0,1)}(\textbf{k}_{1}|\theta) \, \mathcal{K}_{(1,1)}(\textbf{k}_{2} - \textbf{k}_{1}, \textbf{k}_{2}|\theta) \, Q(\textbf{k}_{1}|\theta)    \begin{bmatrix}
        \delta_{r}(\textbf{k}_{2} - \textbf{k}_{1}) & \delta_{i}(\textbf{k}_{2} - \textbf{k}_{1}) \\
        -\delta_{i}(\textbf{k}_{2} - \textbf{k}_{1}) & \delta_{r}(\textbf{k}_{2} - \textbf{k}_{1})
    \end{bmatrix} \nonumber \\
    &+ (\textbf{k}_{2} \leftrightarrow\textbf{k}_{1})\bigg\rbrace \frac{\lambda_{\delta} \, V(\textbf{k}_{1}) \, V(\textbf{k}_{2})}{\mathcal{K}^{2}_{(0,1)}(\textbf{k}_{1}|\theta) \,Q(\textbf{k}_{1}|\theta) \, \mathcal{K}^{2}_{(0,1)}(\textbf{k}_{2}|\theta) \,Q(\textbf{k}_{2}|\theta)} \,,
\end{align}
where, again, the correction to the determinant vanishes at this order.

Using the inverse of the covariance defined above, the likelihood becomes
\begin{align}
    \mathcal{P}^{(1,1)}_{\rm like}[\delta _{o}| \delta, \theta] &\propto \mathcal{P}^{(0,1)}_{\rm like}[\delta _{o}| \delta, \theta] \exp \bigg\lbrace \sum_{\textbf{k}_{1}, \textbf{k}_{2} \in \mathbb{D}} \frac{V(\textbf{k}_{1}) \, V(\textbf{k}_{2})}{\mathcal{K}^{2}_{(0,1)}(\textbf{k}_{1}|\theta) \,Q(\textbf{k}_{1}|\theta) \, \mathcal{K}^{2}_{(0,1)}(\textbf{k}_{2}|\theta) \,Q(\textbf{k}_{2}|\theta)} \nonumber \\
    &\times \Big[ \mathcal{K}_{(0,1)}(\textbf{k}_{1}|\theta) \, \mathcal{K}_{(1,1)}(\textbf{k}_{2} - \textbf{k}_{1}, \textbf{k}_{2}|\theta) \, Q(\textbf{k}_{1}|\theta) \, \delta_{r}(\textbf{k}_{2} - \textbf{k}_{1}) + (\textbf{k}_{2} \leftrightarrow\textbf{k}_{1}) \Big] \nonumber \\
    &\times \Big[ \delta_{o}(\textbf{k}_{1}) - \delta_{t}^{(\leq N, 0)}(\textbf{k}_{1}|\delta, \theta) \Big]\Big[ \bar{\delta} _{o}(\textbf{k}_{2}) - \bar{\delta}_{t}^{(\leq N, 0)}(\textbf{k}_{2}|\delta, \theta) \Big]
    \bigg\rbrace \,.
    \label{eq:like_field_11_finished}
\end{align}

As expected, the likelihood above has terms which is the leading order contribution to the stochastic bispectrum, e.g., $\mathcal{K}_{(0,1)}(\textbf{k}_{1}|\theta) \, \mathcal{K}_{(1,1)}(\textbf{k}_{2} - \textbf{k}_{1}, \textbf{k}_{2}|\theta) \, Q(\textbf{k}_{1}|\theta) \, P_{t}^{(\leq N, 0)}(\textbf{k}|\theta)$. It also contains similar terms, but changing the fully nonlinear power spectrum by the first-order one, which is the leading order theoretical term of the theoretical stochastic bispectrum. 

Note that, despite the fields cubed in the expression, we do not have to evaluate any bispectrum. Actually, this likelihood has quantities like $\left\langle \delta^{3}(\textbf{x})\right\rangle \propto \sum_{\textbf{k}_{1}, \textbf{k}_{2}} \delta(\textbf{k}_{1}) \, \delta(\textbf{k}_{2}) \, \delta(\textbf{k}_{2} - \textbf{k}_{1})$. This makes the likelihood more tractable numerically because there is no need for the computation of the double summations.

The result of this section is the leading order contribution coming from the correlation between the stochastic and deterministic fields. This contrasts with the results of \cite{Cabass2} that have computed the effect of this operator at all orders. The same work also considers the perturbative expansion of their results and gets the same third-order moments [$\left\langle \delta^{3} (\textbf{x}) \right\rangle$] as found here directly by the perturbative approach.

\subsection{\texorpdfstring{$(0,2)$}{02}-term}
\label{ssec:field_02}

Lastly, we consider the leading-order non-Gaussian correction to the stochastic field. We keep the linear term because of its non-perturbative nature but, we do not include the $(1,1)$ term to make easier the interpretation of the final result.

Similarly to Eq.~\eqref{eq:full_field_11}, the \textit{full likelihood} is given by
\begin{align}
    \mathcal{P}^{(0,2)}_{\rm full}[\delta _{o}| \delta, \epsilon,\theta] &=  \int \mathcal{D}\omega_{r} \, \mathcal{D}\omega_{i} \, \exp \bigg\lbrace -i \sum_{\textbf{k} \in \mathbb{D}} \omega (\textbf{k}) \left[ \bar{\delta} _{o}(\textbf{k}) - \bar{\delta}_{t}^{(\leq N, 0)}(\textbf{k}|\delta, \theta) \right] \bigg\rbrace \nonumber \\
    &\times \exp \bigg\lbrace i \,\lambda_{\epsilon} \,\sum_{\textbf{k} \in \mathbb{D}}  \omega (\textbf{k}) \Big[ \,\mathcal{K}_{(0,1)}(\textbf{k}|\theta) \bar{\epsilon}(\textbf{k}) \nonumber \\
    &+ \lambda_{\epsilon} \, \int_{\mathbb{R}^{3}_{\Lambda}} d^{3}q_{1} \, \int_{\mathbb{R}^{3}_{\Lambda}} d^{3}q_{2} \, \mathcal{K}_{(0,2)} (\textbf{q}_{1}, \textbf{q}_{2}|\theta) \, \bar{\epsilon}(\textbf{q}_{1}) \, \bar{\epsilon}(\textbf{q}_{2}) \Big] \bigg\rbrace \,.
\end{align}

Following similar approach as in Sec.~\ref{ssec:computation_second}, we get the following likelihood after marginalizing over the stochastic field
\begin{align}
    \mathcal{P}^{(0,2)}_{\rm like}[\delta _{o}| \delta, \theta] &= \int \mathcal{D}\omega_{r} \, \mathcal{D}\omega_{i} \, \exp \bigg\lbrace -i \sum_{\textbf{k} \in \mathbb{D}} \omega (\textbf{k}) \left[ \bar{\delta} _{o}(\textbf{k}) - \bar{\delta}_{t}^{(\leq N, 0)}(\textbf{k}|\delta, \theta) \right] \bigg\rbrace \nonumber \\
    &\times \exp \Bigg\lbrace i \, \lambda_{\epsilon}^{2} \sum_{\textbf{q}_{1}, \textbf{q}_{2} \in \mathbb{D}} \begin{bmatrix} 
    \omega_{r}(\textbf{q}_{1}) && \omega_{i}(\textbf{q}_{1}) 
    \end{bmatrix} \,
    M(\textbf{q}_{1}, \textbf{q}_{2}|\theta)\begin{bmatrix} 
        \omega_{r}(\textbf{q}_{2}) \\ 
        \omega_{i}(\textbf{q}_{2}) 
    \end{bmatrix} 
    \Bigg\rbrace \,,
\end{align}
where the covariance matrix is defined as
{\small \begin{align}
        M(\textbf{q}_{1}, \textbf{q}_{2}|\theta) &= \mathbb{I} \, \frac{Q(\textbf{q}_{1}|\theta)}{V(\textbf{q}_{1})} \, \delta_{K}(\textbf{q}_{2} - \textbf{q}_{1}) + i \, \lambda_{\epsilon}^{2} \, Q(\textbf{q}_{1}|\theta) \, Q(\textbf{q}_{2}|\theta) \, \frac{\mathcal{K}_{(0,2)}(\textbf{q}_{1}, \textbf{q}_{2}|\theta)}{V(\textbf{q}_{12})} \begin{bmatrix}
        \omega_{r}(\textbf{q}_{12}) & \omega_{i}(\textbf{q}_{12}) \\
        \omega_{i}(\textbf{q}_{12}) & -\omega_{r}(\textbf{q}_{12})
    \end{bmatrix} \,,
\end{align}}
and the Heaviside function is inside the second order kernel.

The likelihood above has the same structure of the one in Eq.~\eqref{eq:like_second}. Therefore, we can immediately read out the final result after marginalizing over the $omega$ field\footnote{Only remember that now the observed density field is shifted by the deterministic theoretical prediction.}.
{ \begin{align}
        \mathcal{P}^{(4)}_{\rm like}[\delta _{o}| \theta] &\approx \mathcal{P}_{\rm like}^{(0)}[\delta _{o}| \theta]  \nonumber \\ 
                                                          &\times \bigg\lbrace 1 +2 \sum_{(\textbf{q}_{1}, \textbf{q}_{2}) \in \mathbb{D}_{2}} \frac{V(\textbf{q}_{1}) \, V(\textbf{q}_{2}) \, Q(\textbf{q}_{1}|\theta) \, \mathcal{K}_{(0,2)}(\textbf{q}_{1}, \textbf{q}_{2}|\theta) \, Q(\textbf{q}_{2}|\theta) \, }{\mathcal{K}_{(0,1)}(\textbf{q}_{1}|\theta) \, \mathcal{K}_{(0,1)}(\textbf{q}_{2}|\theta) \, \mathcal{K}_{(0,1)}^{2}(\textbf{q}_{12}|\theta) \, Q(\textbf{q}_{1}|\theta) \, Q(\textbf{q}_{2}|\theta) \, Q(\textbf{q}_{12}|\theta)}\nonumber \\
                                                          &\times \left[\Delta_{r}(\textbf{q}_{1}) \, \Delta_{r}(\textbf{q}_{2}) \, \Delta_{r}(\textbf{q}_{12}) + 2\Delta_{r}(\textbf{q}_{1}) \, \Delta_{i}(\textbf{q}_{2}) \, \Delta_{i}(\textbf{q}_{12}) - \Delta_{i}(\textbf{q}_{1}) \, \Delta_{i}(\textbf{q}_{2}) \, \Delta_{r}(\textbf{q}_{12})\right] \bigg\rbrace \,,
    \label{eq:like_field_02}
\end{align}}
where 
\begin{equation}
    \Delta_{x}(\textbf{q}) = \delta_{o,x}(\textbf{q}) - \delta_{t,x}^{(\leq N, 0)}(\textbf{q}) \,,
\end{equation}
with $x = r$ or $i$.

Unlike the likelihood of Eq.~\eqref{eq:like_field_11_finished}, the expression above depends on the observed bispectrum, similar to Eq.~\eqref{eq:like_second_final}. 

From the functional expressions of Eq.~\eqref{eq:like_field_11_finished} and Eq.~\eqref{eq:like_field_02} it is possible to see a pattern: the addition of an operator $\propto \delta^{N} \, \epsilon^{M}$ will introduce a dependence in the cross $(N+M+1)$-point function with $N$ theoretical and $M+1$ observed density fields. All other $(N+M+1)$-point functions with less observed density fields will also appear, as expected from the previous sections.

\section{Conclusions}
\label{sec:conclusions}

We used cosmological perturbation theory to compute a self-consistent likelihood for the observed density field conditioned on a set of parameters. The final likelihood automatically gives us:
\begin{itemize}
    \item The optimal summary statistics to use;
    \item The theoretical prediction for all summary statistics;
    \item The functional shape of the likelihood linking both.
\end{itemize}

The final likelihood also constrains the n-point functions and the order in which each one must be computed. It removes the freedom of the standard analysis in choosing both the orders of the predictions and covariances. This more restrictive nature is the price to pay for the consistency of the likelihood.

At the leading order, the likelihood has the same Fisher information as the Gaussian one. It also leads to a very similar numerical posterior, even for small observational volumes. This is in agreement with the central limit theorem. 

At second-order, the likelihood depends on the tree-level bispectrum and the $(2,2)$ one-loop correction of the power spectrum in the same way as a first-order expanded Gaussian. Exact numerical comparisons between the two likelihoods are an interesting topic and will be addressed in future works.

Performing the marginalization only over the stochastic field, it is also possible to compute the field-level likelihoods. In particular, the likelihood including the correlation between the deterministic and stochastic sectors was computed. The final result depends on the leading order contribution to the stochastic bispectrum. Third-order moments of the density field are also present in this case.

The framework presented here is very general and can be applied to any quantity well described by the cosmological perturbation theory. In particular, it might be used to: Study selections of tracers; Get optimal summary statistics for a given observational configuration and order in perturbation theory; Analyze data from galaxy surveys at higher orders, and others.

It also allows the use of different cutoff scales for the data and the initial conditions, as well as different scales in different parts of the data and/or theory.

\section*{Acknowledgments}

The authors thank (in alphabetical order) Henrique Rubira, Jens St\"ucker, and Raul Abramo for the helpful comments on the draft.
RV acknowledges the support from the Juan de la Cierva fellowship (FJC2021-048002-I), the support from project PID2021-128338NB-I00 from the Spanish Ministry of Science, and support from the European Research Executive Agency HORIZON-MSCA-2021-SE-01 Research and Innovation program under the Marie Skłodowska-Curie grant agreement number 101086388 (LACEGAL).

\appendix 

\section{Theoretical model of the tracer's density field}
\label{app:theory}

In the literature, the bias expansion usually takes the form \cite{Desjacques1}
\begin{equation}
    \delta_{t}(\textbf{x}) = \sum_{\mathcal{{O}}} \left[ b_{\mathcal{O}} + \epsilon_{\mathcal{O}}(\textbf{x})\right] \, \mathcal{O}(\textbf{x}) \,,
\end{equation}
we have omitted the dependency in the set of parameters $\theta$, including the theoretical cut-off scale $\Lambda$.

The stochastic fields $\epsilon_{\mathcal{O}}(\textbf{x})$ are completely generic, without any \textit{a priori} constraints in their distributions.

In this work, we choose a different approach where we also expand the stochastic fields in terms of a “primordial” Gaussian stochastic field. Therefore, our theoretical prediction of Eq.~\eqref{eq:theoretical} can also be written as
\begin{equation}
    \delta_{t}(\textbf{x}) = \sum_{\mathcal{{O}}} \left[ b_{\mathcal{O}} + \sum_{\tilde{\mathcal{O}}} b_{\tilde{\mathcal{O}}\tilde{\mathcal{O}}} \, \tilde{\mathcal{O}}(\textbf{x})\right] \, \mathcal{O}(\textbf{x}) \,,
\end{equation}
where the operators $\mathcal{O}(\textbf{k})$ are constructed from local operations with the primordial density field $\delta(\textbf{x})$ and  $\tilde{\mathcal{O}}(\textbf{k})$ from local operations with the “primordial” stochastic field.

Therefore, the kernels $\mathcal{K}_{(n,m)}(\textbf{q}_{1}, \cdots, \textbf{q}_{n}, \textbf{p}_{1}, \cdots, \textbf{p}_{m}|\theta)$ have information about the particular structure of the operator $\mathcal{O}^{(n)}(\textbf{x})\tilde{\mathcal{O}}^{(m)}(\textbf{x})$ including the bias parameters, any cosmological dependence, and the effect of the theoretical cut-off scale.

It is possible to expand the stochastic sector of the theory, similarly to the deterministic part, because the n-point functions of the stochastic field are suppressed by a factor of $\left[P(k|\theta)/Q(k|\theta)\right]^{n-1}$, which is smaller than unity on large scales ($k \lessapprox 0.2 \, [h/{\rm Mpc}]$) and for a mildly dense galaxy sample ($\bar{n}\gtrapprox 10^{-4} \, [{\rm Mpc}/h]^{3}$).

Moreover, in the same way it happens with the deterministic part (as discussed in Sec.~\ref{ssec:computation_second} and App.~\ref{app:covariance}), we are complete up to the (n+1)-point function when considering up to order $n$ operators in the stochastic part. Therefore, when taking the order of the deterministic and stochastic sectors to be the same, we take into account all information of the n-point functions up to this order plus one.

Another advantage of this approach is that we can parametrize any scale dependence in the n-point function of the original stochastic fields $\epsilon_{\mathcal{O}}(\textbf{x})$ through the shape of the kernels using simple free parameters. For instance, the usual $k^{2}$-dependency of the stochastic power spectrum can be parametrized by the operator
\begin{equation}
    \tilde{\mathcal{O}}^{(1)}(\textbf{x}) = (1+c_{0}) \, \epsilon(\textbf{x}) + c_{2} \, R_{\rm NL}^{2} \,\nabla^{2}\epsilon(\textbf{x}) \Rightarrow \mathcal{K}_{(0,1)}(\textbf{k}|\theta) = 1 + c_{0} + c_{2} \,\left( R_{NL} \, k \right)^{2} \,,
\end{equation}
with $\left\langle \epsilon(\textbf{k}) \, \epsilon(\textbf{k}')\right\rangle = (2\pi)^{3} \, \delta_{D}(\textbf{k} + \textbf{k}') \, \frac{1}{\bar{n}}$, and the $c_{0}$ and $c_{2}$ constants being the ones usually considering in the analysis of the power spectrum \cite{Mergulhao1, Mergulhao2, Ivanov1, Nishimichi1}.

Poisson-like stochastic bispectrum is also easily obtained using the operators
\begin{equation}
    \tilde{\mathcal{O}}^{(2)}(\textbf{x}) + \tilde{\mathcal{O}}^{(1)}(\textbf{x})\mathcal{O}^{(1)}(\textbf{x}) = b_{\epsilon^{2}} \, \epsilon^{2}(\textbf{x}) + b_{\delta \epsilon} \, \epsilon(\textbf{x}) \, \delta(\textbf{x})\Rightarrow \left\lbrace \begin{array}{cc}
         \mathcal{K}_{(0,2)}(\textbf{k}_{1}, \textbf{k}_{2}|\theta) = b_{\epsilon^{2}} \\
         \mathcal{K}_{(1,1)}(\textbf{k}_{1}, \textbf{k}_{2}|\theta) = b_{\delta \epsilon}
    \end{array} \right. \,.
\end{equation}

Technically, the expansion of the stochastic sector around a Gaussian field is also helpful as it allows us to handle both fields in the same way, simplifying the computations.

\section{Computation of the full likelihood in Fourier space}
\label{app:like}

All the calculations performed in this work start from the \textit{full likelihood} defined in Eq.~\eqref{eq:full}. In this appendix, we derive its functional from starting from the configuration space fields. For simplicity, in this work, we will assume that the observed field is defined in the whole space. However, taking into account the selection function of a particular galaxy survey is crucial for comparison to data, and will be performed in future work.

Firstly, we define the observed density field of the tracer $a$, smoothed in some cutoff observational scale\footnote{Remember that this scale does not need to be the same as any of the theoretical scales used in the theoretical computations, as discussed in Sec.~\ref{ssec:computation_first}}, as 
\begin{equation}
    \delta^{a}_{o}(\textbf{x}|\Lambda{o}) = \int_{\mathbb{R}^{3}} d^{3}y \, W(\textbf{x - \textbf{y}}|\Lambda{o}) \, \delta^{a}_{o}(\textbf{y}) \,,
\end{equation}
where $\delta^{a}_{o}(\textbf{y})$ is the observed density field of the tracer $a$, and $\Lambda{o}$ is the maximum scale used to analyze the data.

Assuming that we have a reliable theoretical prediction for the density field, as the one in Eq.~\eqref{eq:theoretical}, the \textit{full likelihood} can be written as 
\begin{align}
    \mathcal{P}_{\rm full}[\left\lbrace \delta^{a}_{o} \right\rbrace _{a=1}^{\rm T} | \delta, \left\lbrace \epsilon^{a}  \right\rbrace _{a=1}^{\rm T}, \theta, \left\lbrace \Lambda^{a}_{d}\right\rbrace_{a=1}^{\rm T}] &= \prod _{a=1} ^{\rm T} \left\lbrace \prod _{\textbf{x} \in \mathbb{R}^{3} }\delta _{D} \left[ \delta _{o}^{a}(\textbf{x}|\Lambda^{a}_{d}) - \delta _{t}^{a}(\textbf{x}|\theta) \right] \right\rbrace \nonumber \\
    &= \prod _{a=1} ^{\rm T} \mathcal{P}_{\rm full}[\delta^{a}_{o} | \delta, \epsilon^{a}, \theta, \Lambda^{a}_{d}] \,,
\end{align}
where, in the last line, we defined the \textit{full likelihood} for a single tracer, and we kept open the possibility of smoothing the data at different scales for each tracer, as recommended by \cite{Rubira2} to diminish the effects of fingers-of-god.

Using the Fourier representation of the Dirac delta function and the same definition of path integrals as in the main text, the likelihood of each tracer becomes
\begin{align}
    &\mathcal{P}_{\rm full}[\delta^{a}_{o} | \delta, \epsilon^{a}, \theta, \Lambda^{a}_{d}] = \int \mathcal{D} \omega^{a} \exp \left\lbrace -i \sum _{\textbf{x} \in \mathbb{R}^{3}} \omega^{a}(\textbf{x}) \left[ \delta _{o}^{a}(\textbf{x}|\Lambda{o}) - \delta _{t}^{a}(\textbf{x}|\theta)\right] \right\rbrace \nonumber \\
    &= \int \mathcal{D} \omega^{a} \, \mathcal{D} \bar{\omega}^{a} \, \delta_{D}\left[ \omega(\textbf{x}) - \bar{\omega}(\textbf{x})\right] \exp \left\lbrace -i \sum _{\textbf{x} \in \mathbb{R}^{3}} \omega^{a}(\textbf{x}) \left[ \delta _{o}^{a}(\textbf{x}|\Lambda{o}) - \delta _{t}^{a}(\textbf{x}|\theta)\right] \right\rbrace \nonumber \\
    &\propto \int \mathcal{D} \omega^{a}_{r} \, \mathcal{D} \omega^{a}_{i} \, \delta_{D}\left[ \omega_{r}(\textbf{k}) - \omega_{r}(-\textbf{x})\right] \, \delta_{D}\left[ \omega_{i}(\textbf{k}) + \omega_{i}(-\textbf{x})\right] \nonumber \\
    &\times \exp \left\lbrace -i \left[ \sum _{\textbf{k} \in \mathbb{R}_{\Lambda{o}}^{3}} \omega^{a}(\textbf{k}) \, \bar{\delta}^{a}_{o}(\textbf{k}) - \sum _{\textbf{k} \in \mathbb{R}^{3}} \omega^{a}(\textbf{k}) \, \bar{\delta}^{a}_{t}(\textbf{k}|\theta)  \right] \right\rbrace
    \label{eq:full_like_k}
\end{align}
where, from the first to the second line, we have introduced the complex conjugate of the real field $\omega(\textbf{x})$ together with a gauge-fixing-like term to impose the reality of the field, and in the third line we have used Parseval's theorem to write the integral in Fourier space. In the last line, we also made the change of (field) variables:
\begin{align}
    \omega_{r}(\textbf{k}) &= \frac{V(\textbf{k})}{2} \, \sum _{\textbf{x} \in \mathbb{R}^{3}} \left[  e^{i \textbf{k} \cdot \textbf{x}} \, \omega(\textbf{x}) + e^{-i \textbf{k} \cdot \textbf{x}} \, \bar{\omega}(\textbf{x}) \right] \nonumber \\
    \omega_{i}(\textbf{k}) &= \frac{V(\textbf{k})}{2i} \, \sum _{\textbf{x} \in \mathbb{R}^{3}} \left[  e^{i \textbf{k} \cdot \textbf{x}} \, \omega(\textbf{x}) - e^{-i \textbf{k} \cdot \textbf{x}} \, \bar{\omega}(\textbf{x}) \right]\,,
\end{align}
with Jacobian 
\begin{equation}
    \left| \frac{\partial \left[ \omega_{r}(\textbf{k}), \omega_{i}(\textbf{k}) \right]}{\partial \left[ \omega(\textbf{x}), \bar{\omega}(\textbf{x}) \right]}\right| = \frac{V^{2}(\textbf{k})}{4} \,\left| \begin{bmatrix}
        e ^{i\textbf{k} \cdot \textbf{x}} && e ^{-i\textbf{k} \cdot \textbf{x}} \\
        -ie ^{i\textbf{k} \cdot \textbf{x}} && ie ^{-i\textbf{k} \cdot \textbf{x}}
    \end{bmatrix}\right| = i\frac{V^{2}(\textbf{k})}{2} ,.
\end{equation}

Note that the first sum over $\mathbb{R}^{3}$ was truncated to $|\textbf{k}|\leq \Lambda{o}$ due to the definition used for the observed filtered density field.

The second sum in the exponential of the last line of Eq.~\eqref{eq:full_like_k} can be broken into a term with $|\textbf{k}| \leq \Lambda{o}$ and another one for the smaller scales. Then, the integral over the modes with $|\textbf{k}| > \Lambda{o}$ can be performed separately 
\begin{align}
    &\int \mathcal{D} \omega^{a}_{r} \, \mathcal{D} \omega^{a}_{i} \, \delta_{D}\left[ \omega_{r}(\textbf{k}) - \omega_{r}(-\textbf{x})\right] \, \delta_{D}\left[ \omega_{i}(\textbf{k}) + \omega_{i}(-\textbf{x})\right] \exp \left[ i \sum _{\textbf{k} \in \mathbb{R}^{3}} \omega^{a}(\textbf{k}) \, \bar{\delta}^{a}_{t}(\textbf{k}|\theta)  \right] \nonumber \\
    &= \prod_{\textbf{k} \in \bar{\mathbb{R}}^{3}_{\Lambda{o}}/\mathbb{Z}_{2}} \delta_{D}\left[ \delta_{t}^{a}(\textbf{k}|\theta)\right]
\end{align}
where $\bar{\mathbb{R}}^{3}_{\Lambda{o}} = \left\lbrace \textbf{k} \in \mathbb{R}^{3} \, | \, \textbf{k}^{2} > \Lambda{o}^{2}\right\rbrace$.

This term only enforces that our theoretical prediction must be zero for all modes larger than $\Lambda{o}$, which can be easily accomplished using the same filtering in the data and the theoretical prediction. 

The remaining sums in Eq.~\eqref{eq:full_like_k} can be split into the positive and negative parts of some (arbitrary) direction
{\small \begin{align}
    \sum _{\textbf{k} \in \mathbb{R}_{\Lambda{o}}^{3}} \omega^{a}(\textbf{k}) \left[  \bar{\delta}^{a}_{o}(\textbf{k})   - \bar{\delta}^{a}_{t}(\textbf{k}|\theta) \right] &= \sum _{\textbf{k} \in \mathbb{R}_{\Lambda{o}}^{3}/\mathbb{Z}_{2}} \left\lbrace \omega^{a}(\textbf{k}) \left[ \bar{\delta}^{a}_{o}(\textbf{k})   - \bar{\delta}^{a}_{t}(\textbf{k}|\theta) \right] + \omega^{a}(-\textbf{k}) \left[ \delta^{a}_{o}(\textbf{k})   - \delta^{a}_{t}(\textbf{k}|\theta) \right] \right\rbrace \nonumber \\
    &=2 \sum _{\textbf{k} \in \mathbb{R}_{\Lambda{o}}^{3}/\mathbb{Z}_{2}} \omega^{a}(\textbf{k}) \left[  \bar{\delta}^{a}_{o}(\textbf{k})   - \bar{\delta}^{a}_{t}(\textbf{k}|\theta) \right] \,,
\end{align}}
where, in the first line, we have used the reality of $\delta_{t}^{a}(\textbf{x}|\theta)$, and in the second line we performed the integrals over the negative part using the Dirac delta functions together with the fact that $\omega^{a}(\textbf{k}) \, \bar{\delta}^{a}(\textbf{k}|\theta)$ is real.

Plugging the result above back into Eq.~\eqref{eq:full_like_k}, and redefining the fields $\omega_{r}$ and $\omega_{i}$, we get
\begin{align}
    \mathcal{P}_{\rm full}[\delta^{a}_{o} | \delta, \epsilon^{a}, \theta, \Lambda^{a}_{d}] &= \int \mathcal{D} \omega^{a}_{r} \, \mathcal{D} \omega^{a}_{i}\exp \left\lbrace -i \sum _{\textbf{k} \in \mathbb{R}_{\Lambda{o}}^{3}/\mathbb{Z}_{2}} \omega^{a}(\textbf{k}) \left[ \, \bar{\delta}^{a}_{o}(\textbf{k}) -\bar{\delta}^{a}_{t}(\textbf{k}|\theta)  \right] \right\rbrace \nonumber \\
    &= \prod_{\textbf{k}\in \mathbb{R}_{\Lambda{o}}^{3}/\mathbb{Z}_{2}} \delta_{D}[\delta^{a}_{o}(\textbf{k}) -\delta^{a}_{t}(\textbf{k}|\theta)] \,,
\end{align}
which is the same likelihood used in the main text and defined in Eq.~\eqref{eq:full}.

The interpretation of this result is straightforward: we have to consider only half of $\mathbb{R}^{3}_{\Lambda{o}}$ because of the reality condition of the density fields. Considering the whole space would double-count information.

Despite the different spaces of integration, it is not hard to show that the likelihood derived above and the one using the entire space are related by a square root
\begin{equation}
    \mathcal{P}_{\rm full}[\delta^{a}_{o} | \delta, \epsilon^{a}, \theta, \Lambda^{a}_{d}] = \prod_{\textbf{k}\in \mathbb{R}_{\Lambda{o}}^{3}/\mathbb{Z}_{2}} \delta_{D}[\delta^{a}_{o}(\textbf{k}) -\delta^{a}_{t}(\textbf{k}|\theta)] = \sqrt{\prod_{\textbf{k}\in \mathbb{R}_{\Lambda{o}}^{3}} \delta_{D}[\delta^{a}_{o}(\textbf{k}) -\delta^{a}_{t}(\textbf{k}|\theta)]} \,.
\end{equation}

Therefore, it is trivial to relate the results of this work with the ones coming from the EFT-likelihood of \cite{Cabass1, Schmidt1}\footnote{Even after considering the marginalization of some fields, the relationship still holds because all integrals can be split into the positive and negative parts of $\mathbb{R}^{3}_{\Lambda{o}}$.}.

\section{Non-linear covariance matrix}
\label{app:covariance}

In this appendix, we analyze the non-linear covariance matrix first presented in Eq.~\eqref{eq:second_like_unfinished}. We compute its inverse and determinant up to fourth order on $\lambda$. In the main text, we use only terms up to order $\lambda^{2}$, but the $\lambda^{4}$ terms help in the interpretation of the likelihood at higher orders.

This covariance is defined as the object multiplying the square of $\Phi(\textbf{k})$ in Eq.~\eqref{eq:second_like_unfinished}
{\small \begin{align}
    M^{-1}(\textbf{q}_{1}, \textbf{q}_{2}|\theta) &=  V(\textbf{q}_{1}) \, V(\textbf{q}_{2}) \left[ \tilde{C}^{-1}(\textbf{q}_{1}|\theta) \frac{\delta_{K}(\textbf{q}_{2} - \textbf{q}_{1})}{V(\textbf{q}_{1})} -i \frac{\lambda^{2}\theta_{H}(\Lambda_{o} - |\textbf{q}_{12}|)}{V(\textbf{q}_{12})} \tilde{\mathcal{K}}_{2}(\textbf{q}_{1}, \textbf{q}_{2}|\textbf{q}_{12}, \theta) \right] \,,
    \label{eq:nonlinear_covariance}
\end{align}}
where both $\tilde{C}(\textbf{q}_{1}|\theta)$ and $\tilde{\mathcal{K}}_{2}(\textbf{q}_{1}, \textbf{q}_{2}|\textbf{k}, \theta)$ are matrices defined in Eq.~\eqref{eq:Phi}. 

Firstly, we want to find the inverse matrix defined by
\begin{equation}
    \sum_{\textbf{q}_{2} \in \mathbb{R}^{3}_{\Lambda}}  M(\textbf{q}_{1}, \textbf{q}_{2}) \, M^{-1}(\textbf{q}_{2}, \textbf{q}_{3}) = \delta _{K}(\textbf{q}_{3} - \textbf{q}_{1}) \,,
    \label{eq:inverse}
\end{equation}
using the perturbative ansatz
{ \begin{equation}
    M(\textbf{q}_{1}, \textbf{q}_{2}| \theta) = \frac{\tilde{C}(\textbf{q}_{1}|\theta)}{V(\textbf{q}_{1})} \left[\lambda ^{0} \, M^{(0)}(\textbf{q}_{1}, \textbf{q}_{2}|\theta) \, + \, \lambda ^{2} \, M^{(2)}(\textbf{q}_{1}, \textbf{q}_{2}|\theta) \, + \, \lambda ^{4} \, M^{(4)}(\textbf{q}_{1}, \textbf{q}_{2}|\theta) + \cdots \right]\,.
\end{equation}}

Plugging the ansatz above into Eq.~\eqref{eq:inverse} and using the definition of the nonlinear inverse covariance [Eq.~\eqref{eq:nonlinear_covariance}], we get the following matrices that add up to make the inverse
{ \begin{align}
    M^{(0)}(\textbf{q}_{1}, \textbf{q}_{3}|\theta) &= \mathbb{I} \, \delta _{K}(\textbf{q}_{3} - \textbf{q}_{1}) \,;  \\ 
    M^{(2)}(\textbf{q}_{1}, \textbf{q}_{3}|\theta) &= \frac{i \, V(\textbf{q}_{1})}{V(\textbf{q}_{13})} \, \theta_{H}(\Lambda_{o} - |\textbf{q}_{13}|) \,\tilde{\mathcal{K}}_{2}(\textbf{q}_{1}, \textbf{q}_{3}|\textbf{q}_{13}, \theta) \, \tilde{C}(\textbf{q}_{3}|\theta) \,; \\
    M^{(4)}(\textbf{q}_{1}, \textbf{q}_{3}|\theta) &= -\sum_{\textbf{q}_{2} \in \mathbb{R}^{3}_{\lambda}} \frac{V(\textbf{q}_{1})V(\textbf{q}_{2})}{V(\textbf{q}_{12})V(\textbf{q}_{23})} \left[ \tilde{\mathcal{K}}_{2}(\textbf{q}_{1}, \textbf{q}_{2}|\textbf{q}_{12}, \theta) \, \tilde{C}(\textbf{q}_{2}|\theta) \, \tilde{\mathcal{K}}_{2}(\textbf{q}_{2}, \textbf{q}_{3}|\textbf{q}_{23}, \theta) \right] \nonumber \\
    &\times \tilde{C}(\textbf{q}_{3}|\theta) \, \theta_{H}(\Lambda_{o} - |\textbf{q}_{12}|) \, \theta_{H}(\Lambda_{o} - |\textbf{q}_{23}|) \,,
\end{align}}
where $\mathbb{I}$ is the $4\times4$ identity matrix.

From the expressions above, we can see that the $M^{(0)}$ term gives the expression we got for the first-order likelihood in Eq.~\eqref{eq:like_first}, as expected. The term $M^{(2)}$ will generate an $\omega^{3}$ term that will integrate to the tree-level bispectrum. Lastly, the $M^{(4)}$ term will integrate to something proportional to $\left\langle |\mathcal{O}^{(2)}(\textbf{k})|^{2}\right\rangle$, which is one of the one-loop corrections to the power spectrum. Terms proportional to the trispectrum will never be generated, even considering infinite order in the inversion of the covariance. We need to start with the $\lambda^{3}$ terms in the perturbative expansion to be able to generate four-point functions. This finding is directly aligned with the results of \cite{Schmidt3}, which shows that a nth-order perturbative expansion in the density field only captures information up to the (n+1)-point function, even when higher-order terms are kept while computing the likelihood.

In addition to the nonlinear covariance, we also need to compute its determinant\footnote{This coincides with the usual definition of the determinant by a factor of $(2\pi)^{N}$, where $N$ is formally infinity.}
\begin{equation}
    \sqrt{{\rm det}_{\Lambda} \left[ M\right](\theta)} \coloneqq \int \mathcal{D}\Phi \, \exp \left\lbrace - \frac{1}{2} \sum_{\textbf{q}_{1}, \textbf{q}_{2} \in\mathbb{R}^{3}_{\Lambda}} \Phi_{a}(\textbf{q}_{1}) \, \left[ M^{-1}(\textbf{q}_{1}, \textbf{q}_{2}|\theta)\right]^{ab} \, \Phi_{b}(\textbf{q}_{2})\right\rbrace \,.
\end{equation}

For the computation of the determinant perturbatively, it is also convenient to introduce the trace as
\begin{equation}
    {\rm Tr}_{\Lambda}  \left[ M \right](\theta) = \sum _{i,j=1}^{4} \sum_{\textbf{q}_{1}, \textbf{q}_{2} \in\mathbb{R}^{3}_{\Lambda}} M_{ij}(\textbf{q}_{1}, \textbf{q}_{2}|\theta) \, \delta _{K} ^{ij} \, \delta _{K} (\textbf{q}_{2} - \textbf{q}_{1}) \,,
\end{equation}
where $\delta _{K}$ is the usual Kronecker delta symbol for discrete variables.

As done before, we will assume a perturbative ansatz for the determinant
{\small \begin{equation}
    {\rm det}_{\Lambda} \left[ M ^{-1}\right](\theta) = \left\lbrace \prod _{\textbf{k} \in \mathbb{R}^{3}_{\Lambda}}  \,{\rm det} \left[ V(\textbf{k})\,  \tilde{C}^{-1}(\textbf{k}|\theta) \right]\right\rbrace \, \left[ \lambda ^{0} \, D^{(0)}(\theta) \, + \, \lambda ^{2} \, D^{(2)}(\theta) \, + \, \lambda ^{4} \, D^{(4)}(\theta) \, + \cdots\right] \,,
\end{equation}}

Note that the determinant in the product is a usual one taken over $4\times4$ matrices and the index $\textbf{k}$ is explicitly shown to make it clear.

To compute the determinant perturbatively, we use the relationship
\begin{equation}
    {\rm det}_{\Lambda} \left[ M \right](\theta) = e^{\ln {\rm det}_{\Lambda} \left[ M \right](\theta)} = e^{{\rm Tr}_{\Lambda} \left[ \ln M \right](\theta)} \,.
    \label{eq:det}
\end{equation}

Using our ansatz in the expression above, and expanding the exponential and logarithm in Taylor, we get the following expression for each order of the determinant 
{ \begin{align}
    D^{(0)}(\theta) &= 1\,; \\
    D^{(2)}(\theta) &= -i \sum_{\textbf{q} \in \mathbb{R}^{3}_{\Lambda}} \frac{V(\textbf{q}) }{V(2\textbf{q})}\, {\rm Tr} \left[ \tilde{C}(\textbf{q}|\theta) \, \tilde{\mathcal{K}}_{2}(\textbf{q}, \textbf{q}|2\textbf{q}, \theta) \right] \, \theta_{H}(\Lambda_{o} - 2|\textbf{q}|) \,; \\
    D^{(4)}(\theta) &= -\frac{1}{2}  \left\lbrace \sum_{\textbf{q} \in \mathbb{R}^{3}_{\Lambda}} \frac{V(\textbf{q}) }{V(2\textbf{q})}\, {\rm Tr} \left[ \tilde{C}(\textbf{q}|\theta) \, \tilde{\mathcal{K}}_{2}(\textbf{q}, \textbf{q}|2\textbf{q}, \theta) \right]  \theta_{H}(\Lambda_{o} - 2|\textbf{q}|) \right\rbrace ^{2} \nonumber \\
    & +\frac{1}{2} \sum_{\textbf{q}_{1}, \textbf{q}_{2} \in \mathbb{R}^{3}_{\Lambda}} \frac{V(\textbf{q}_{1} )V(\textbf{q}_{2})}{V^{2}(\textbf{q}_{12})} \, \theta_{H}(\Lambda_{o} - |\textbf{q}_{12}|) \nonumber \\
    &\times {\rm Tr} \left[ \tilde{C}(\textbf{q}_{1}|\theta) \, \tilde{\mathcal{K}}_{2}(\textbf{q}_{1}, \textbf{q}_{2}|\textbf{q}_{12}, \theta) \, \tilde{C}(\textbf{q}_{2}|\theta) \, \tilde{\mathcal{K}}_{2}(\textbf{q}_{2}, \textbf{q}_{1}|\textbf{q}_{12}, \theta) \right]   \,.
\end{align}}

From Eq.~\eqref{eq:Phi}, we see that $\tilde{C}(\textbf{k}|\theta) \propto \mathbb{I}$ and ${\rm Tr}\left[ \tilde{\mathcal{K}}_{2}(\textbf{q}_{1}, \textbf{q}_{2}|\textbf{q}_{12}|\theta)\right] = 0$. Therefore, the terms with ${\rm Tr} \left[ \tilde{C}(\textbf{q}|\theta) \, \tilde{\mathcal{K}}_{2}(\textbf{q}, \textbf{q}|2\textbf{q}, \theta) \right]$, in $D^{(2)}$ and $D^{(4)}$, are zero. It makes the first correction to the first-order determinant to appear only at order $\lambda^{4}$.

\bibliographystyle{JHEP}
\bibliography{main}

\end{document}